\documentclass[twocolumn]{aastex631}
\usepackage{graphicx}
\usepackage{natbib}
\usepackage{longtable}
\usepackage{rotating}
\usepackage{appendix}
\newcommand\aastex{AAS\TeX}

\received{2021 September 9}
\revised{2021 November 2}
\accepted{2021 November 15}
\submitjournal{ApJ}

\shorttitle{\aastex\ Frequency Curves}
\shortauthors{Lee et al. 2021}

\begin{document}
\defcitealias{2019Lee}{Paper I}
\defcitealias{2020Lee}{Paper II}
\defcitealias{2019Schmidt}{Sch19}

\title{Properties of Fast and Slow Bars Classified by Epicyclic Frequency Curves from Photometry of Barred Galaxies}

\correspondingauthor{Ho Seong Hwang}
\email{galaxy79@snu.ac.kr}

\author[0000-0002-0786-7307]{Yun Hee Lee}
\affil{Korea Astronomy and Space Science Institute (KASI), 776 Daedeokdae-ro, Yuseong-gu, Daejeon 34055, Korea}

\author{Myeong-Gu Park}
\affil{Department of Astronomy and Atmospheric sciences, Kyungpook National University, Daegu, 41566, Korea}

\author{Ho Seong Hwang}
\affil{Astronomy Program, Department of Physics and Astronomy, Seoul National University, 1 Gwanak-ro, Gwanak-gu, Seoul 08826, Korea}
\affil{SNU Astronomy Research Center, Seoul National University, 1 Gwanak-ro, Gwanak-gu, Seoul 08826, Korea}
\affil{Korea Astronomy and Space Science Institute (KASI), 776 Daedeokdae-ro, Yuseong-gu, Daejeon 34055, Korea}

\author{Hong Bae Ann}
\affiliation{Department of Earth Science Education, Pusan National University, Busan, 46241, Korea}

\author{Haeun Chung}
\affil{University of Arizona, Steward Observatory, 933 N Cherry Ave, Tucson, AZ 85721, USA}

\author{Taehyun Kim}
\affil{Department of Astronomy and Atmospheric sciences, Kyungpook National University, Daegu, 41566, Korea}

\begin{abstract}
We test the idea that bar pattern speeds decrease with
time owing to angular momentum exchange with a dark matter halo. If this
process actually occurs, then the locations of the corotation resonance and
other resonances should generally increase with time. We therefore derive the angular velocity $\Omega$ and epicyclic frequency $\kappa$ as
functions of galactocentric radius for 85 barred galaxies using photometric
data. Mass maps are constructed by assuming a dynamical mass-to-light ratio
and then solving the Poisson equation for the gravitatonal potential. The
location of Lindblad resonances and the corotation resonance radius are
then derived using the standard precession frequency curves in conjunction
with bar pattern speeds recently estimated from the Tremaine-Weinberg method as
applied to Integral Field Spectroscopy (IFS) data. Correlations between
physical properties of bars and their host galaxies indicate that bar
{\it length} and the corotation radius depend on the disk circular velocity
while bar {\it strength} and pattern speed do not. As the bar pattern speed
decreases, bar strength, length, and corotation radius incease, but when
bars are subclassified into fast, medium, and slow domains, no significant
change in bar length is found. Only a hint of an increase of bar strength
from fast to slow bars is found. These results suggest that bar length in
galaxies undergoes little evolution, being instead determined mainly by
the size of their host galaxy.

\end{abstract}

\section{Introduction}\label{chap1}
Bars in galaxies can be described by three properties: bar length, strength, and pattern speed \citep{2015Aguerri}. Numerical simulations suggest that these parameters will change over time, owing to the angular momentum exchange between the bar and the dark halo. When the bar is deprived of its angular momentum by the dark halo, the pattern speed of the bar slows down, the corotation radius moves outward, and the bar grows in length and strength in sequence \citep{1980Sellwood, 1985Weinberg, 2000Debattista, 2002Athanassoula, 2003Athanassoula, 2014Athanassoula}. In particular, the pattern speed of the bar $\Omega_{\rm bar}$ constrains the dynamics of a disk galaxy by determining the locations of the corotation and Lindblad resonances \citep{1987Binney}.  

The resonances predicted in the density wave theory \citep{1964Lin} have provided an effective theoretical basis for understanding disk galaxies. For example, spiral density waves can propagate between the inner Lindblad resonance (ILR) and the outer Lindblad resonance (OLR) \citep{1989Adams, 1996Bertin}. In the presence of weak ovals or low contrast bars, test-particle
simulations \citep{1981Schwarz, 1984aSchwarz, 1980Simkin} have shown
that nuclear, inner, and outer rings secularly develop near the principal
resonances: ILR, the 4:1 (Ultraharmonic) resonance (UHR), and the OLR,
respectively. In the presence of a strong bar, the concept of an ILR may
not exist, and formation of a nuclear ring will depend on the presence of
the $x_2$ orbit family \citep{2004Regan}. The lengths of nuclear bar and of large-scale bar are correlated to the ILR and corotation radius (CR), respectively \citep{1999Rautiainen}. It means that the pattern speed of a bar determines the sizes of rings and bars. 

To measure the pattern speed, we need kinematic information from spectroscopy, while the bar length and strength can be calculated from photometric images. Although many indirect ways to measure the bar pattern speed from photometry have been proposed \citep{1979Roberts, 1983Prendergast, 1997Puerari, 2008Rautiainen, 2009Buta, 2012Perez, 2017Buta}, the most reliable way is the Tremaine-Weinberg method \citep[hereafter TW method]{1984Tremaine} that measures the bar pattern speed from spectroscopy directly. The pattern speed is derived from the mean line-of-sight (LOS) velocity over several positions, based on the continuity equation. The measurement requires time-consuming long-silt observations with several positions parallel to the line of nodes \citep{1995Merrifield}. However, the recent Integral Field Spectroscopy (IFS) data facilitates the measurement of the bar pattern speed by making it possible to obtain multiple pseudo long-slits from a single observation. The Calar Alto Legacy Integral Field Area (CALIFA) \citep{2012Sanchez}, the Mapping Nearby Galaxies at APO (MaNGA) \citep{2015Bundy}, and the Multi Unit Spectroscopic Explorer (MUSE) surveys have been used to measure the bar pattern speeds for $\sim$ 100 galaxies at $0 < z < 0.15$ so far \citep{2015Aguerri, 2019Guo, 2019Cuomo, 2020Garma-Oehmichen, 2021Williams}. 

When the pattern speed of a bar is known, the corotation radius
$R_{\rm CR}$ can be estimated from the rotation curve. While the IFS data allows
rotation curves to be derived, a reasonable assumption that can be made is
that the rotation curve is flat in the corotation region \citep{2015Aguerri, 2019Guo, 2019Cuomo}. With an estimate of the bar radius, $R_{\rm bar}$, we are then able to
derive the important ratio $\cal R = R_{\rm CR}/R_{\rm bar}$ from infrared images. \citet{2000Debattista} suggested that in a minimum halo, a bar would be limited to $\cal R \rm = 1.4$. The limit to how far a bar can extend is given by the orbital calculations of \citet{1980Contopoulos}. He showed that stellar orbits are aligned parallel to the bar supporting the shape of the bar inside the corotation radius, but they are perpendicular to the bar beyond it. Therefor, the ratio has been used to classify barred galaxies into slow $(\cal R \rm > 1.4)$, fast $(1 \le \cal R \rm \le 1.4)$ and ultrafast $(\cal R \rm < 1)$  bars \citep{2003Aguerri, 2015Aguerri, 2019Guo, 2019Cuomo, 2020Garma-Oehmichen}.



However, previous observations have not yet found any coherent clues for bar evolution \citep{2012Perez, 2015Aguerri, 2019Guo, 2020Cuomo, 2021Kim}; most galaxies are in the phase of a fast bar \citep{2015Aguerri, 2019Cuomo, 2020Cuomo}, which could imply the dark halos have low concentration \citep{2000Debattista} or inefficient angular momentum exchange \citep{2002Athanassoula, 2013Athanassoula}. On the other hand, simulations including the Evolution and Assembly of GaLaxies and their Environments (EAGLE) \citep{2015Schaye, 2016McAlpine} and the Illustris The Next Generation (IllustrisTNG) \citep{2018Nelson, 2019Nelson} showed that most barred galaxies have slow bars \citep{2017Algorry, 2021Roshan}. When it comes to host galaxies, observations do not show any significant correlation between the ratio $\cal R$ and galactic properties, which include morphological type, stellar mass, dark matter, age, and metallicity \citep{2015Aguerri, 2019Guo, 2020Garma-Oehmichen, 2020Cuomo}. \citet{2012Perez} explored the evolution of $\cal R$ at $z < 0.8$ but found no change of $\cal R$ with redshift. \citet{2021Kim} also showed little or no evolution in the bar length at $0.2 < z \le 0.835$ from the Cosmological Evolution Survey with Hubble Space Telescope (HST/COSMOS).

In this paper, we try a different approach to study the bar evolution by examining the bar pattern speed on the frequency curves of $\Omega-\kappa/2$, $\Omega-\kappa/4$, $\Omega$, and $\Omega+\kappa/2$. Each of these curves decreases with increasing galactocentric
radius, such that if the bar pattern speed decreases with time, the radius
of corotation, the locations of the inner and outer Lindblad resonances
(ILR, OLR) and ultraharmonic resonance (inner 4:1 resonance, or UHR) all
increase with time. To construct frequency curves, \citet{2019Schmidt} estimated a potential profile from velocity curves by assuming an axisymmetric Miyamoto-Nagai gravitational potential \citep{1975Miyamoto}. \citet{2020Garma-Oehmichen} obtained angular velocity curves from spectroscopic data by fitting the VELFIT model \citep{2007Spekkens}. In this work, we derive frequency curves from photometry. We construct the mass map by applying the dynamical mass-to-light ratio \citep{2015vandeSande} to the surface brightness distribution and analyze the potential map constructed by solving the Poisson equation  \citep{2001Buta, 2020Lee}. We utilize the bar pattern speed measured by the TW method from spectroscopy in the literature \citep{2015Aguerri, 2019Guo, 2019Cuomo, 2020Garma-Oehmichen}. 



This paper is organized as follows. Section \ref{chap2} introduces our sample obtained from Pan-STARRs DR1 data archive. Section \ref{chap3} describes the processes where we obtain frequency curves from photometry. Section \ref{chap4} shows the results on the relation between the bar pattern speed and the frequency curves. We classify barred galaxies into fast, medium, and slow bars that might be related to the bar evolution. We compare the classifications with properties of host galaxies and bars. Sections \ref{chap5} and \ref{chap6} are assigned to discussion and summary, respectively.

\section{Samples and Data} \label{chap2}
We collect sample galaxies whose bar pattern speeds $\Omega_{\rm bar}$ are measured from recent IFS observations including CALIFA and MaNGA with the TW method \citep{2015Aguerri, 2019Guo, 2019Cuomo, 2020Garma-Oehmichen}. There are 89 galaxies in total (apart from duplication). We obtain their optical images from the Pan-STARRs DR1 data archive (PS1). The sample galaxies are distributed from SB0 to SBc for their Hubble type. The TW method was first designed for early-type barred galaxies where a large fraction of old stars are assumed to obey the continuity equation \citep{1984Tremaine, 2011Corsini}, but has been reliably applied to various conditions, including late-type spirals or gas tracer (H$\alpha$) \citep{2006Emsellem, 2009Fathi, 2015Aguerri, 2019Guo, 2019Cuomo, 2020Garma-Oehmichen}. The CALIFA sample galaxies are distributed in $0.005 < z < 0.03$ and $-19.5 \le M_{\rm r} \le -22.5$ for the redshift and absolute SDSS $r-$band magnitude, respectively \citep{2020Cuomo}. The MaNGA sample galaxies span in the range of $0.02 < z < 0.15$ and $-19.5 \le M_{\rm r} \le -23$ \citep{2020Cuomo}.

In addition, we analyze PS1 images of IC 1438 and NGC 2835  whose frequency curves were derived by \citet{2019Schmidt} to compare the results from this analysis on photometry with those measured from spectroscopy. \citet{2019Schmidt} derived their frequency curves from the potential by fitting the rotation curves along with three other galaxies. They measured the corotation radius using two photometric estimations: Fourier analysis of azimuthal profile \citep{1997Puerari} and the change of the dust lane \citep{1979Roberts, 1983Prendergast}. They determined ILR, UHR, CR, and OLR by comparing the pattern speed with the frequency curves.

For the 91 galaxies, we collect \textit{g\textsubscript{P1}} and \textit{i\textsubscript{P1}} band images from the PS1. The Pan-STARRs with Gigapixel Camera, mounted at Haleakala Observatories on the island of Maui, Hawaii, provides a good image quality with pixel scale of $0.\arcsec258$, and FWHM of $1.\arcsec31$ and $1.\arcsec11$ for \textit{g\textsubscript{P1}} and \textit{i\textsubscript{P1}}, respectively \citep{2020bMagnier}. We deconvolve the images to obtain sharper rotation curves in the central region removing a seeing effect by applying the Lucy-Richardson algorithm using the FWHM of each band \citep{2020Chung}, which influences the measurement of the ILRs. 

We mask foreground stars, adjacent galaxies, and stellar clumpy regions within the target galaxies for automatic analyses \citep{2019Lee}. We deproject galaxies using the orientation parameters, position angle and ellipticity, and reject five galaxies that are highly elongated due to the high inclinations of $\sim$70 degrees. Because the resulting bar properties including the bar length, strength, and pattern speed are very sensitive to the orientation parameter \citep{2019Zou, 2020Lee, 2020Garma-Oehmichen}, we use the orientation parameters reported in the literature \citep{2015Aguerri, 2019Guo, 2019Cuomo, 2020Garma-Oehmichen} to make a fair comparison. We also reject one galaxy with the pattern speed of nearly zero, $\Omega_{\rm bar}=0.4~\rm km~s^{-1}~kpc^{-1}$. Accordingly, there are 85 barred galaxies in the final sample. We present the parameters including inclination, position angle (PA), and bar pattern speed we used in Appendix. 

\section{Calculation of Frequency}\label{chap3}
\subsection{Frequency Curves from Surface Brightness}\label{chap3.1}
Stellar orbits in weak non-axisymmetric potentials can be described by the epicycle theory of nearly circular orbits in an axisymmetric potential \citep{1987Binney}. The circular orbit frequency, namely angular velocity $\Omega(r)$, is derived from the gravitational potential $\Phi (x,y)$ as follows,

\begin{eqnarray}\label{Eq4.1a}
\Omega^2(r)=\left<\frac{1}{r}\frac{\partial\Phi}{\partial r}\right>.
\end{eqnarray}
The epicyclic frequency $\kappa(r)$ with which stars move inward and outward in the circular motion is determined as
\begin{eqnarray}\label{Eq4.1b}
\kappa^2(r)=\left<\frac{\partial^2\Phi}{\partial x^2}+\frac{\partial^2\Phi}{\partial y^2}+2\left(\frac{1}{r}\frac{\partial\Phi}{\partial r}\right)^2\right>
\end{eqnarray}
where $\left< \right>$ stands for an azimuthal average  \citep{1987Binney, 1990Pfenniger, 2006Michel, 2019Schmidt}.
The corotation radius is defined as the radius where the stellar angular velocity becomes the same as the bar pattern speed, $\Omega = \Omega_{\rm bar}$. Resonances occur when the difference between the stellar angular velocity and the pattern speed multiplied by an integer becomes the epicyclic frequency: $m(\Omega-\Omega_{\rm bar}) = \pm \kappa$ for integer values of $m$ \citep{1987Binney, 1998Elmegreen}. Although we cannot be sure whether the stellar orbits become resonant when the axisymmetry is broken by a strong bar, the epicyclic approximation has been widely used to estimate resonance locations \citep{1981Schwarz, 1993Combes, 1994Byrd, 1996Buta, 1996Combes, 1999Buta, 2002Buta, 2006Michel, 2019Schmidt, 2021Williams}.

The gravitational potential can be constructed with the assumption of a constant mass-to-light ratio (\textit{M/L}) \citep{1994Quillen}. The photometric surface brightness distribution is translated into the mass distribution, which yields the two-dimensional potential through the Poisson equation. The bar strength is the ratio of the transverse force to the radial one (force ratio, hereafter) and can be calculated from the potential as well \citep{2001Buta, 2020Lee}. 

\subsubsection{From Light To Mass} \label{chap3.1.1}
The determination of $M/L$ allows us to translate the photometric luminosity to the stellar mass. \citet{2001Bell} first explored the relation between the $M/L$ and the color by comparing the observed colors with the stellar population synthesis (SPS) model. \citet{2015vandeSande} developed the relation into the dynamical $M/L$ by the direct stellar kinematic mass measurements,  estimated from the effective radius, S$\acute{e}$rsic index, and velocity dispersion measurements \citep{2006Cappellari}. This does not depend on any assumptions, including the metallicity, stellar initial mass function (IMF), and the SPS models. 

We construct our mass maps from the color and the absolute magnitude using the following formula: 
\begin{eqnarray}\label{eq3.1.1b}
\rm log_{10}\frac{\it M_{\rm dyn}}{\it M_{\odot}} = -1.27 + 1.75(\it g-i) - \rm 0.4(\it M_i-\it M_{i,\odot}),
\end{eqnarray}
which is derived from the relation between the dynamical $M/L$ ratio and the color \citep[see Table 3]{2015vandeSande} by comparing the absolute magnitude of the Sun. The dynamical mass within each pixel is determined by its $i-$band absolute magnitude from $i_{P1}$ band images and the $g-i$ color of the galaxy. We adopt the mean $g-i$ color within a scale length $h_{r}$ in the radial color profile as the $g-i$ color of a galaxy. We use the PS1 $i$-band solar absolute magnitude of $M_{\rm i, \odot} = 4.52$ \citep[in AB,][]{2018Willmer}. However, we note that the relation in \citet{2015vandeSande} was explored for massive quiescent galaxies with a mass limit of $M_{\ast} > 10^{11} M_{\odot}$ and with a color selection of $U-V > (V-J)\times0.88+0.59$ \citep{2009Williams}.


\subsubsection{Calculation of Potential Map} \label{chap3.1.2} 
The potential map was constructed following the procedures of \citet{2020Lee} by solving the Poisson equation with the Fast Fourier Transform (FFT) on Cartesian coordinates \citep{1969Hohl, 1994Quillen, 2001Buta}. In constructing the potential map, the vertical density distribution is assumed to follow the exponential model \citep{2002Laurikainen, 2004Buta, 2020Lee}. Two-dimensional mass maps are converted to three-dimensional mass distribution by convolving it with the vertical density profile. The vertical scale height is taken from the ratio of the disk scale length and vertical scale height, $h_r/h_z$, considering the different disk thicknesses according to the Hubble types $T$: 4 for $T \le 1$, 5 for $2 \le T \le 4$, and 9 for $T \ge 5$ \citep{1998deGrijs, 2004bLaurikainen, 2016Diaz, 2020Lee}. We measure the scale length $h_r$ with the exponential fit to the surface brightness profile at $i-$band, obtained from the IDL-based ellipse fitting \citep{2019Lee}.
\subsubsection{Frequecy Curves}\label{chap3.1.3}
\begin{figure*}[htb]
\includegraphics[bb = 310 240 610 500,  width = \linewidth, clip =]{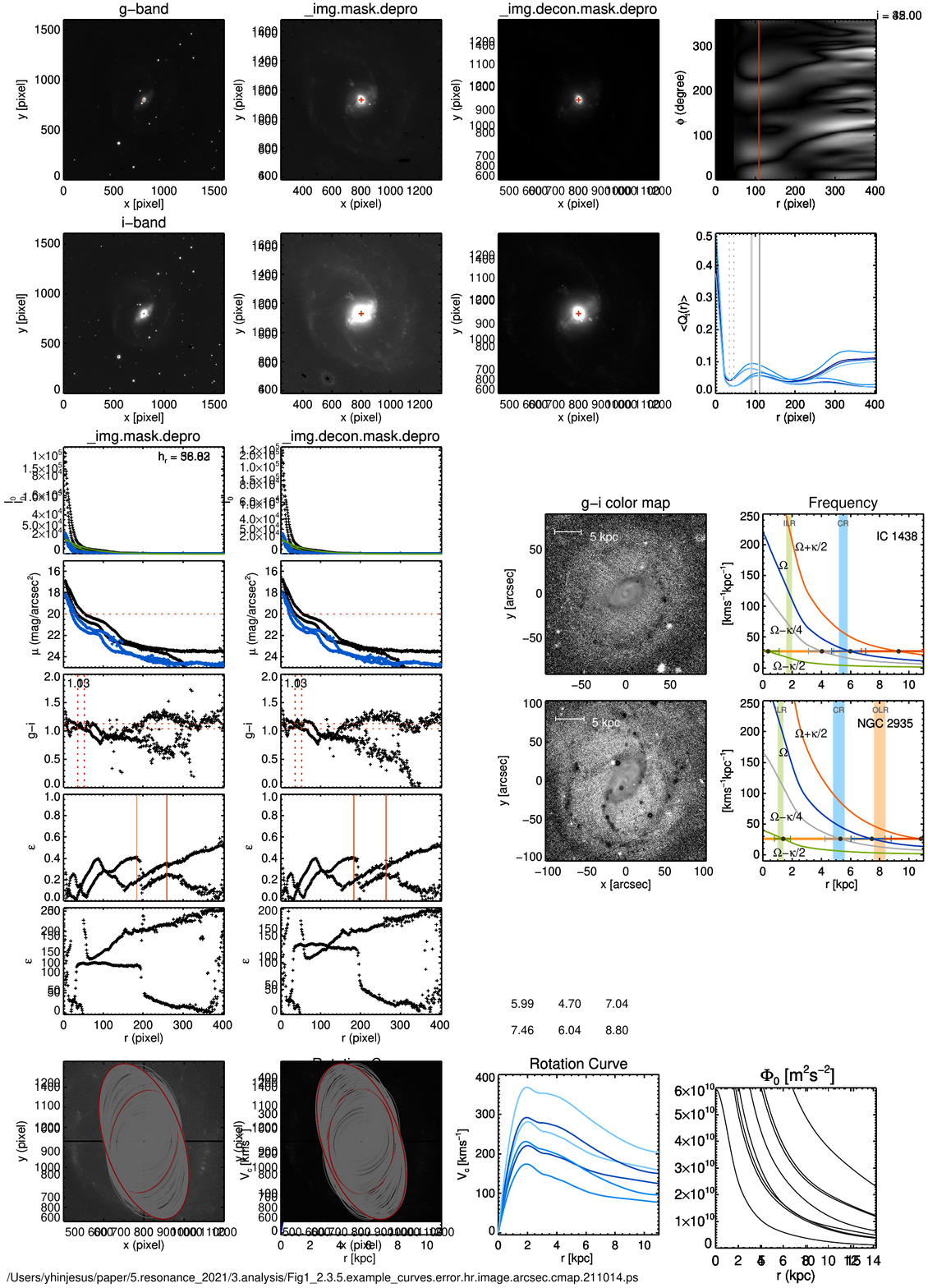}
\caption{Examples of $g-i$ color maps (left) and frequency curves (right) for IC 1438 (top row) and NGC 2935 (bottom row) from the potential map based on photometry. In the left panels, the nuclear, inner, and outer rings appear bluer in color than their surroundings. In the right panels, the green, gray, blue and red curves show $\Omega - \kappa/2$, $\Omega - \kappa/4$, $\Omega$, and $\Omega + \kappa/2$, in sequence. The orange horizontal line indicates the bar pattern speed $\Omega_{\rm bar}$ from the literature \citep{2019Schmidt}. The solid black circles show the intersecting points of the frequency curves and the bar pattern speed. They are $R_{\rm ILR}$, $R_{\rm UHR}$, $R_{\rm CR}$, and $R_{\rm OLR}$, in sequence, from the center. The horizontal error bar represents the uncertainty of each resonance location.} The green, sky blue, and orange columns from top to bottom, respectively, represent the $R_{\rm ILR}$, $R_{\rm CR}$, and $R_{\rm OLR}$ with error ranges calculated by \citet{2019Schmidt}. \label{Fig1}
\end{figure*}

Figure \ref{Fig1} shows examples of the frequency curves (right panels) for IC 1438 and NGC 2935 obtained from our photometric approach with our adopted dynamical $M/L$ ratio. We display $\Omega - \kappa/2$, $\Omega - \kappa/4$, $\Omega$, and $\Omega + \kappa/2$ by green, gray, blue, and red curves, respectively. The bar pattern speed $\Omega_{\rm bar}$ determines the corotation radius where $\Omega_{\rm bar}$ intersects the circular orbit frequency $\Omega(r)$. In the same way, the locations of the ILR, UHR, and OLR are determined by $\Omega_{\rm bar} = \Omega - \kappa/2$, $\Omega_{\rm bar} = \Omega - \kappa/4$, and $\Omega_{\rm bar} = \Omega + \kappa/2$, respectively \citep{1987Binney}. The pattern speeds $\Omega_{\rm bar}$ of these two galaxies were estimated by \citet{2019Schmidt} from photometric methods mentioned above. $\Omega_{\rm bar}$ is displayed by a horizontal line of orange color. $R_{\rm ILR}$ (green), $R_{\rm UHR}$ (gray), $R_{\rm CR}$ (blue), and $R_{\rm OLR}$ (red) are presented with uncertainties on the horizontal line of $\Omega_{\rm bar}$.

We estimate the uncertainties from (1) the scatter of the relation between the dynamical $M/L$ and the color \citep{2015vandeSande} and (2) the difference between the assumptions of a thick ($h_r/h_z=4$) and thin ($h_r/h_z=9$) disk. The scatter of $\rm log \it M_{\rm dyn}/\it L$ is larger than the orthogonal scatter of the best-fitting line by a factor of 1.5 \citep[see Figure 3]{2015vandeSande}. Therefore, we measure the uncertainties of $\rm log \it M_{\rm dyn}/\it L_i$ by multiplying by 1.5 the mean orthogonal scatter of the best fitting (i.e., 1.3) in $i$-band. The green, sky blue, and orange columns from top to bottom indicate the $R_{\rm ILR}$, $R_{\rm CR}$, and $R_{\rm OLR}$ with error ranges calculated by \citet{2019Schmidt}.

\subsection{Comparison with the results in the Literature} \label{chap3.2} 
\citet{2019Schmidt} calculated the frequency curves for five spiral galaxies including IC 1438 and NGC 2935. They constructed the radial velocity curves from the H$\alpha$ emission line observations and calculated the gravitational potential by fitting the rotational curves to the axisymmetric Miyamoto-Nagai gravitational potential, 
\begin{eqnarray}\label{Eq4.2}
\Phi(R,z) = -\frac{GM}{\sqrt{R^2+[a+\sqrt{z^2+b^2}]}^2}, 
\end{eqnarray}
where $\Phi(R,z)$ is the Miyamoto-Nagai potential at $(R,z)$ and $M$ indicates the total mass \citep{1975Miyamoto}. The parameters $a$ and $b$ are shape parameters, which represent a flattened disk distribution with the ratio $b/a\sim0.4$, an ellipsoidal distribution with $b/a=1$, and a spherical distribution with $b/a\sim5$ \citep{1987Binney, 2019Schmidt}. They estimated the potential considering two or three components designated by $b/a$ and calculated the angular velocity with Equation \ref{Eq4.1a}. They used the equation for epicyclic frequency $\kappa(r)$ as
\begin{equation}
\kappa^2=4\Omega^2[1+\frac{1}{2}\left(\frac{R}{\Omega}\frac{d\Omega}{dR}\right)]  
\end{equation}
\citep{1998Elmegreen}.

Figure \ref{Fig1} shows the galaxies in common between \citet{2019Schmidt} and this study. For IC 1438 (top row), our measurements of $R_{\rm CR}$ and $R_{\rm OLR}$ are consistent with the estimates of \citet[see Figure 3]{2019Schmidt} within errors, while $R_{\rm ILR}$ is located inward compared to theirs. When comparing $R_{\rm ILR}$, we note different ways to deal with a bulge: they adopted an ellipsoidal distribution of $b/a=1$, while we considered the bulge region to be exponentially distributed in the z direction like the disk. In the case of NGC 2935 (bottom row), our estimate of $R_{\rm ILR}$ is similar to that of \citet{2019Schmidt}, whereas $R_{\rm CR}$ and $R_{\rm OLR}$ are larger than theirs (but still similar considering the uncertainty).

The right panels of Figure \ref{Fig1} show that IC 1438 and NGC 2935 have all kinds of resonances, including ILR, UHR, CR, and OLR. In the left panels, their color index maps show blue features appearing the circumnuclear, inner, and outer rings. It is interesting that three kinds of rings are located near ILR, UHR, and OLR, in sequence \citep{2019Schmidt}. Outer rings are occasionally associated with the outer 4:1 resonance located between CR and OLR according to their subclass, $\rm R_1$ or $\rm R'_1$ \citep{2017Buta}.

\begin{figure}[tb]
\includegraphics[bb = 160 480 300 620, width = \linewidth, clip =]{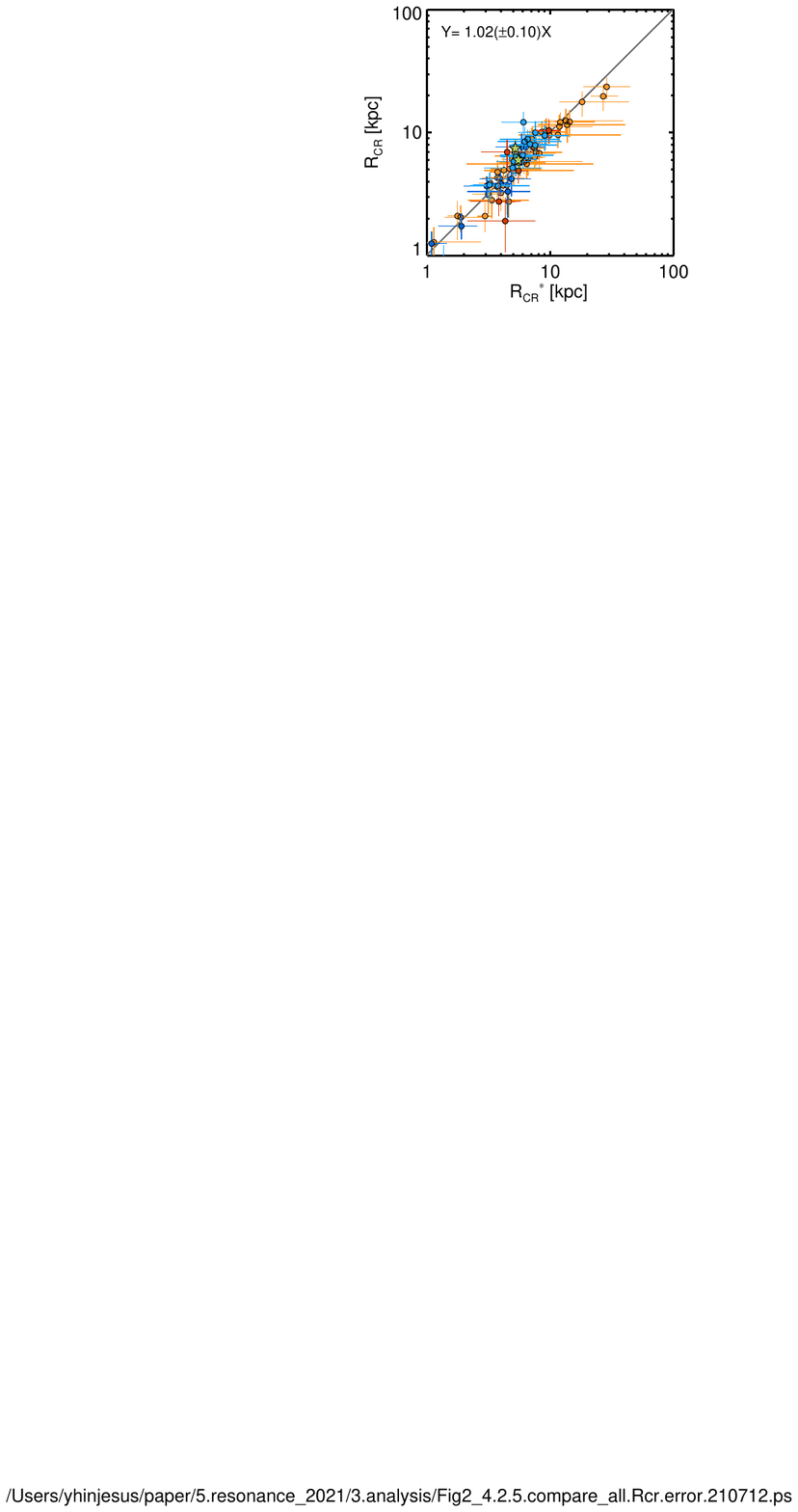}
\caption{Comparison of the corotation radius between our estimations ($R_{\rm CR}$) and those in the literature ($R_{\rm CR}^{\ast}$). The different colors denote galaxies analyzed from different studies: sky blue for \citet{2015Aguerri}, blue for \citet{2019Cuomo}, orange for \citet{2019Guo}, and red for \citet{2020Garma-Oehmichen}. The two green stars represent IC 1438 and NGC 2935 from \citet{2019Schmidt}. The linear fit is represented by a solid line and shown at the top left.}\label{Fig2}
\end{figure}

To provide the result for a larger sample, we present Figure \ref{Fig2} showing the comparison of our estimated corotation radii ($R_{\rm CR}$) and those in the literature obtained from spectroscopy (${R_{\rm CR}}^{\ast}$) \citep{2015Aguerri, 2019Schmidt, 2019Guo, 2019Cuomo, 2020Garma-Oehmichen}. As noted in Section \ref{chap2}, the bar pattern speeds $\Omega_{\rm bar}$ as determined from the TW method \citep{1984Tremaine} are taken from the literature, except for IC 1438 and NGC 2935 (green filled stars). The different colors indicate different literature. \citet{2020Garma-Oehmichen} (red circle) modelled the angular velocity by VELFIT, while the rest \citep{2015Aguerri, 2019Guo, 2019Cuomo} estimated the corotation radius from $R_{\rm CR}=V_{\rm flat}/\Omega_{\rm bar}$ by assuming a flat rotation. We find that our measurement of $R_{\rm CR}$ agrees with $R_{\rm CR}^\ast$ regardless of the variety of ways employed in the literature.

\section{Results} \label{chap4}
\subsection{Bar Properties by Pattern Speed}\label{chap4.1}
\subsubsection{Corotation Radius versus Bar Length} \label{chap4.1.1} 
\begin{figure*}[htb]
\includegraphics[bb = 25 645 465 780,  width = \linewidth, clip =]{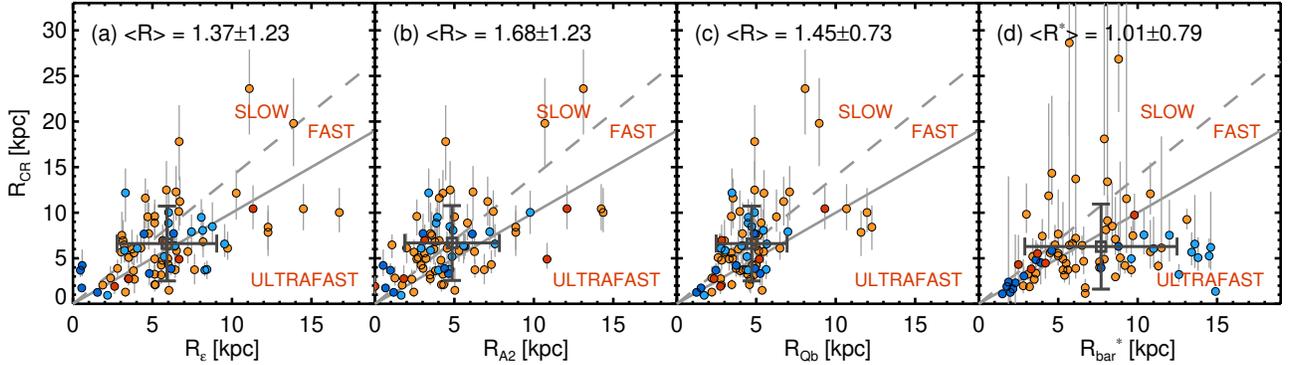}
\caption{Relation between the bar length $R_{\rm bar}$ and the corotation radius $R_{\rm CR}$. $R_{\rm CR}$ is estimated from the frequency curve and the bar pattern speed. The bar length is measured by (a) ellipse fitting ($R_{\epsilon}$), (b) Fourier analysis ($R_{\rm A2}$), and (c) force ratio ($R_{\rm Qb}$). (d) $R_{\rm CR}$ and $R_{\rm bar}^\ast$ are obtained from the literature. The dashed ($R_{\rm CR} = 1.4 R_{\rm bar}$) and solid ($R_{\rm CR} = R_{\rm bar}$) lines denote the criteria for slow, fast, and ultrafast bars. We present the mean and standard deviation of the bar length and the corotation radius by black square with error bars. The mean and standard deviation of $\cal R$ are presented at the top left in each panel. For a comparison with previous results, we use color coded circles indicating different sources of the samples, sky blue for \citet{2015Aguerri}, blue for \citet{2019Cuomo}, orange for \citet{2019Guo}, and red for \citet{2020Garma-Oehmichen}.
}\label{Fig3}
\end{figure*}


As an indicator of the bar pattern speed, the distance-independent ratio $\cal R \rm =R_{\rm CR}/R_{\rm bar}$ has been used in hydrodynamical simulations to model gas and shocks \citep{1996aLindblad, 1996bLindblad, 2001Weiner}. \citet{2000Debattista} suggested the limit of $\cal R \rm \le 1.4$ for a fast bar that ends its slowdown in a maximum disk (minimum halo). The upper limit to where a bar can extend was suggested as $\cal R \rm =1$ in orbital calculations \citep{1980Contopoulos}. On this basis, observational studies have classified barred galaxies into slow ($\cal R \rm >1.4$), fast ($1 < \cal R \rm \le 1.4$), and ultrafast ($\cal R \rm <1$) bars \citep{2003Aguerri, 2015Aguerri, 2019Guo, 2019Cuomo, 2020Garma-Oehmichen}.

The bar length has been usually estimated by ellipse fitting \citep{1995Martin, 1995Wozniak, 2004Jogee} or Fourier analysis \citep{1990Ohta, 2002Laurikainen, 2002Lauri_Salo}. Because bars do not end with sharp edges, it is not trivial to define the full length. Although there have been lots of efforts to find the best way to determine the length of a bar, each method has its own strengths and weaknesses \citep{2002Atha_Mis, 2006Michel, 2021Cuomo}. The widely used way is to measure the radius where the ellipticity, $\epsilon$, or the normalized Fourier amplitude, $A_2$, profile reaches a maximum, at $R_{\epsilon}$ or $R_{\rm A_2}$ even though it cannot estimate the full length of the bar \citep{1995Wozniak, 2002Atha_Mis, 2002Laurikainen, 2002Lauri_Salo}. Similarly, we can define the bar radius where the radial profile of the transverse-to-radial force ratio has a plateau or a maximum peak, $R_{\rm Q_b}$ \citep{2020Lee, 2021Cuomo}.

Figure \ref{Fig3} shows the relation between $R_{\rm CR}$ and $R_{\rm bar}$ measured from $R_{\epsilon}$, $R_{\rm A2}$, and $R_{\rm Qb}$, together with the regimes of slow, fast, and ultrafast bars classified by the ratio $\cal R$. The dashed and solid lines represent $R_{\rm CR} = 1.4 R_{\rm bar}$ and $R_{\rm CR} = R_{\rm bar}$, respectively. We present the mean value and standard deviation of the $R_{\rm bar}$ and $R_{\rm CR}$ by black square and error bars. The mean value of $\cal R$ is given with the standard deviation at the top left in each panel. In panel (d), $R_{\rm CR}$ and $R_{\rm bar}^\ast$ are adopted from the literature \citep{2015Aguerri, 2019Guo, 2019Cuomo, 2020Garma-Oehmichen}.

Basically, the classification by $\cal R$ depends on the method to measure the bar length. In our measurement, galaxies are categorized into 29 slow bars, 9 fast bars, and 40 ultrafast bars when $R_{\epsilon}$ is adopted for $R_{\rm bar}$ (Figure \ref{Fig3}(a)). Five galaxies are rejected because they have no $R_{\rm CR}$ intersected by $\Omega_{\rm bar}$ in their angular velocity curves because of their rapid pattern speeds. Ultrafast bars with $\cal R \rm \le 1$ cannot exist theoretically, but they have existed in observations \citep{2019Cuomo, 2021Cuomo, 2019Guo}. The possibility of real ultrafast bars was raised also by \citet{2007Zhang}.
When we use $R_{\rm A2}$ (Figure \ref{Fig3}(b)) or $R_{\rm Qb}$ (Figure \ref{Fig3}(c)) instead, the number of slow bars increases to 39 galaxies, and the number of fast bars increases to 15 or 13 galaxies. The number of ultrafast bars decreases to 24 or 26 galaxies. On the other hand, when we adopt the values directly from the literature (Figure \ref{Fig3}(d)), sample galaxies are classified into 20 slow bars, 23 fast bars, and 40 ultrafast bars. Except for the sample of \citet{2019Guo} (orange circle), most galaxies are contained within the regimes of the fast and ultrafast bars. 

The mean values of $\cal R$ are 1.37, 1.68, and 1.45, respectively for $R_{\epsilon}$, $R_{\rm A2}$, and $R_{\rm Qb}$. They are somewhat larger compared to $\cal R \rm =1.01\pm0.79$ in the literature due to the smaller bar length in our measurements. The mean bar length is $7.7$ kpc in the literature, while it is measured to be $5.9, 4.9, 4.7$ kpc by $R_{\epsilon}$, $R_{\rm A2}$, and $R_{\rm Qb}$, respectively. We obtain the smallest $\cal R$ from $R_{\epsilon}$ and the largest $\cal R$ from $R_{\rm A2}$. This can more or less reconcile the difference of $\cal R$ between observations and simulations.  Observations have preferentially used the ellipse fitting method to measure the bar length \citep{2015Aguerri, 2019Guo, 2019Cuomo, 2020Garma-Oehmichen}, whereas simulations have utilized Fourier analysis \citep{1996aLindblad, 1996bLindblad, 2000Debattista}. Although observations often examined the bar length from the bar-to-interbar intensity ratio based on the Fourier analysis \citep{2015Aguerri, 2019Guo, 2019Cuomo}, it is different from the way that the bar length is usually calculated with the simulations \citep{1996aLindblad, 1996bLindblad, 2000Debattista}. We will discuss the issues for the bar length measurements and the bar evolution on $\cal R$ in Section \ref{chap5.1} and \ref{chap5.2}, respectively.



\subsubsection{Bar Pattern Speed versus Frequency curves} \label{chap4.1.2} 
\begin{figure*}[htb]
\includegraphics[bb = 20 415 520 790,  width = \linewidth, clip =]{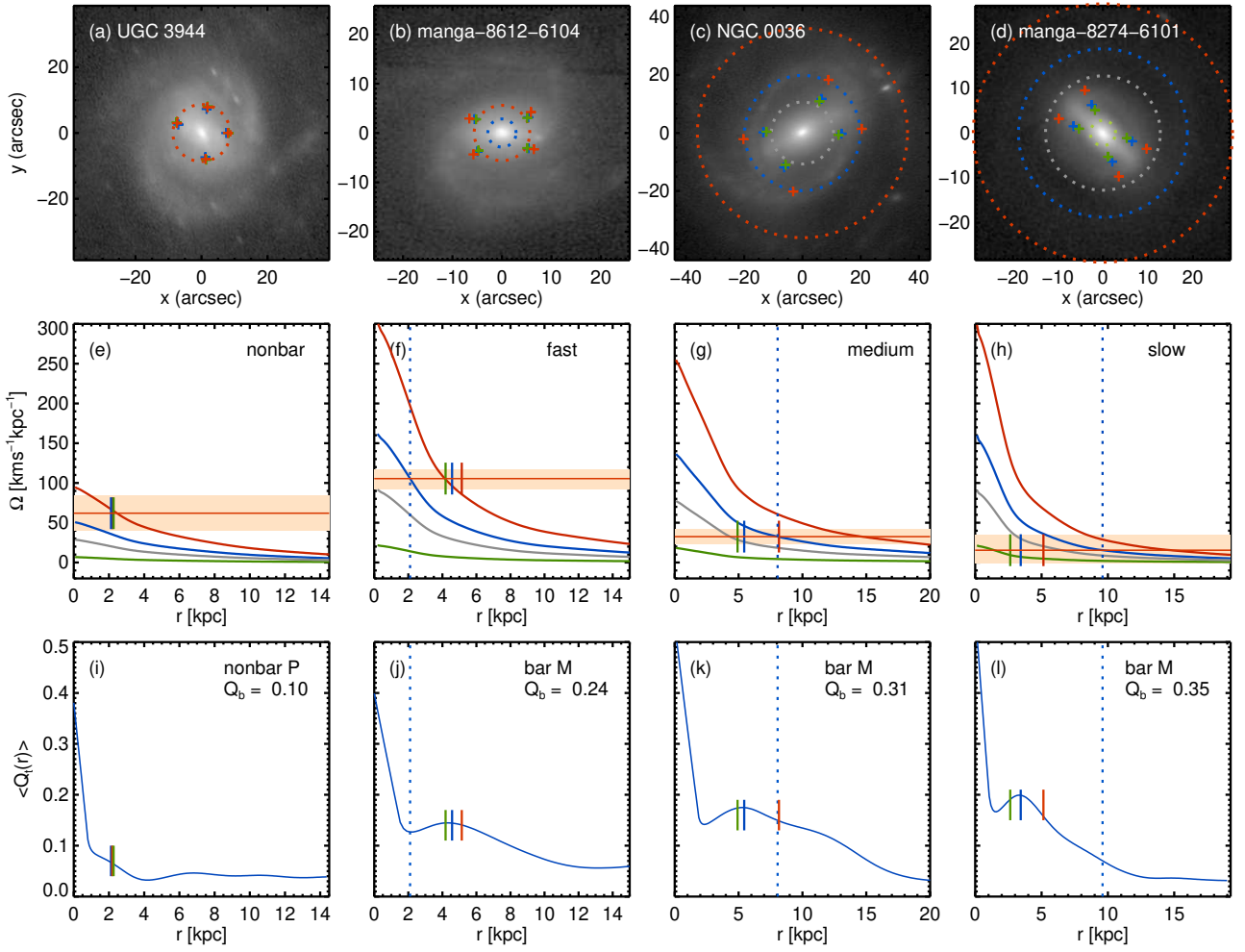}
\caption{Example galaxies for (a) a nonbar, (b) a fast bar, (c) a medium bar, and (d) a slow bar with {\it i}-band deprojected images on the log scale (top row), frequency curves (middle row), and radial profiles of transverse-to-radial force ratio (bottom row). In the top row, we placed circles indicating OLR, CR, UHR, and ILR with red, blue, gray, and green dotted lines. Red, green, and blue crosses indicate bar positions at $R_{\rm bar}$ measured by ellipse fitting ($R_{\epsilon}$), and Fourier analysis ($R_{\rm A2}$), and force ratio ($R_{\rm Qb}$), respectively. The bar positions are determined by analyzing force ratio maps from $R_{\rm bar}$ measured by each method \citep{2020Lee, 2021Cuomo}. The middle row shows the relation between the bar pattern speed and the frequency curves for each class. The red, blue, gray, and green curves represent $\Omega+\kappa/2$, $\Omega$, $\Omega-\kappa/4$, and $\Omega-\kappa/2$, respectively: OLR only for the nonbarred galaxy (e), CR and OLR for the fast bar (f), UHR, CR, and OLR for the medium bar (g), and ILR, UHR, CR, and OLR for the slow bar (h). The blue dotted vertical line displays the corotation radius, and short vertical sticks represent the bar length measured by ellipse fitting (red), Fourier analysis (green), and force ratio (blue). The bottom row shows the radial profile of the transverse-to-radial force ratio. The bar classification based on the force ratio map and the bar strength are organized at the top right. When the radial profile has a plateau, it is classified as type P (i). Type M indicates a galaxy with a maximum peak on the radial profile (j)-(l).} \label{Fig4}
\end{figure*}

\begin{figure*}[htb]
\includegraphics[bb = 10 645 400 780,  width = \linewidth, clip =]{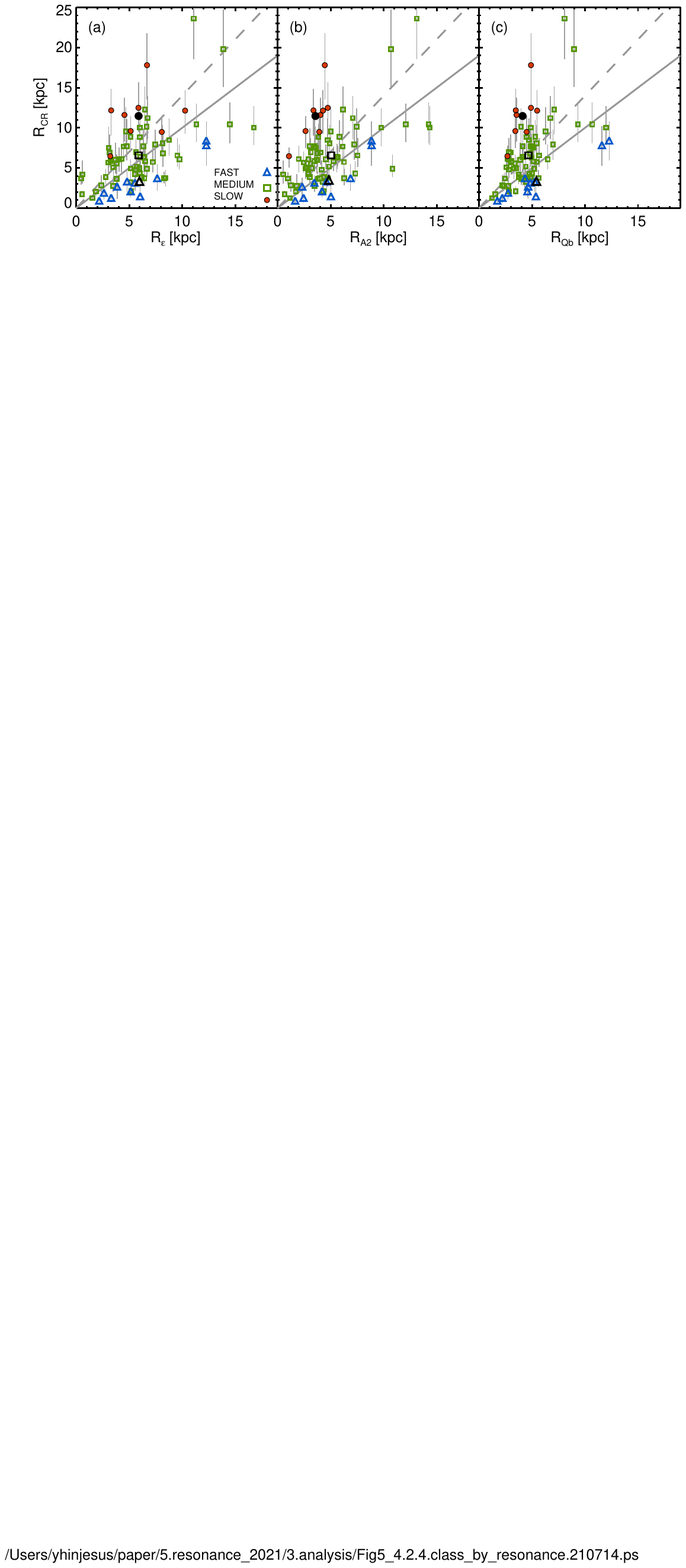}
\caption{Relation between the corotation radius versus the bar length measured by (a) ellipse fitting, (b) Fourier analysis, and (c) force ratio. Newly defined fast, medium, and slow bars \citep{1994Byrd} are represented by blue triangles, green squares, and solid red circles. The fast, medium, and slow bars are roughly placed in the regions of ultrafast, fast, and slow bars, respectively, classified by ratio $\cal R$. The black symbols indicate the mean values of the bar length and the corotation radius for each class.} \label{Fig5}
\end{figure*}

Here, we take a different approach to utilize the bar pattern speed more effectively. That is to compare the bar pattern speed with the Lindblad precession frequency curves directly. If the bar pattern speed decreases by the exchange of angular momentum between a bar and a dark halo \citep{2002Athanassoula, 2003Athanassoula, 2000Debattista}, then the locations of all resonances except the iILR will increase
with time. It will cross more frequency curves from $\Omega$ to $\Omega-\kappa/4$ or $\Omega-\kappa/2$, in turn, producing a corotation radius (CR), an Ultraharmonic Resonance (UHR), and one or two Inner Lindblad Resonances (ILRs). Using these resonances, \citet{1994Byrd} suggested a more physical classification to define a fast, a medium, and a slow bar when the bar pattern speed sets up CR, UHR, and ILR in sequence. Although we cannot trace the process of the slowdown of the bar for a given galaxy, we can glean "snapshots" for galaxies with fast, medium, and slow bars from observational data.

Figure \ref{Fig4} displays example galaxies with images (top row) for each class. The second row shows the frequency curves with the pattern speed for a fast (f), a medium (g), and a slow bar (h). The horizontal lines represent the bar pattern speed with uncertainties. The red, blue, gray, and green curves describe $\Omega + \kappa/2$, $\Omega$, $\Omega - \kappa/4$, and $\Omega - \kappa/2$, respectively. We present the corotation radius by a dotted vertical line. Although we indicate the bar length, $R_{\epsilon}$ (red), $R_{\rm A2}$ (green), and $R_{\rm Qb}$ (blue) by vertical sticks together, this classification is irrelevant to the bar length, which is different from the classification based on the ratio $\cal R$. In the bottom row, we display the radial profiles of force ratio for each galaxy, as introduced in \citet{2020Lee}. It helps to understand the evolution of bars, which we will discuss in Section \ref{chap5.3}.

Among 83 galaxies, we find five galaxies with higher bar pattern speed, not intersecting the stellar angular velocity curve $\Omega$ (Figure \ref{Fig4}(e)). All of them are also classified as nonbarred galaxies from the analysis of the ratio map \citep{2020Lee}. According to the definition of \citet{1994Byrd}, a fast bar is the one hosting a CR and an OLR because of a high pattern speed crossing $\Omega$ and $\Omega+\kappa/2$ (Figure \ref{Fig4}(f)). A medium bar in a galaxy is defined with UHR, CR, and OLR when the pattern speed crosses the frequency curves of $\Omega-\kappa/4$, $\Omega$, and $\Omega+\kappa/2$ (Figure \ref{Fig4}(g)). A slow bar has all kinds of resonances (Figure \ref{Fig4}(h)). \citet{1994Byrd} used the simulations and found a hint in evolutionary stages for fast, medium, and slow bars. For example, a fast bar shows an outer ring, while a medium bar has outer and inner rings; a slow bar shows a nuclear ring as well as outer and inner rings. 

According to our classification scheme, there are eight slow bars, 59 medium bars, and 11 fast bars. Five galaxies without a CR are categorized into nonbarred galaxies. \citet{1982vanAlbada} showed models with one ILR or two ILRs when an $x_2$ family extends to the center or stops before the center. However, we do not find any galaxies with two ILRs in our sample, probably because the data we analyze are not good enough to resolve the central region within $\sim1$ kpc from the center where inner ILR is likely to be located. We also note that the measured ILRs in this study appear slightly smaller than those in \citet{2019Schmidt}, which might be caused by different ways to deal with a bulge in the potential calculation (even though the difference is not larger considering the uncertainty). There could be some slow bars missed in this study for a similar reason.

Figure \ref{Fig5} shows the relation between the bar length and the corotation radius. This plot is similar to Figure \ref{Fig3}, but galaxies here are classified into fast (blue triangle), medium (green square), and slow bars (solid red circle) according to the relation between the bar pattern speed and the frequency curve. The newly defined fast, medium, and slow bars are placed in a similar sequence of ultrafast, fast, and slow bars classified by the ratio $\cal R$, though they are not perfectly corresponding to each other. The newly defined fast bars fall in the region occupied by the ultrafast bars defined by $\cal R$. 

We present the mean values of $\cal R$ for newly defined fast, medium, and slow bars in Table \ref{Table4.1.2}. Although there are some differences according to the measurements of the bar length, they increase from 0.62 to 1.48 and 2.82, in average, from fast to medium and slow bars. We intend to investigate the properties of barred galaxies in terms of the newly defined classes of the fast, medium, and slow bar in Section \ref{chap4.2} and \ref{chap4.3}. 

\begin{deluxetable}{l|ccc}
 \tablecaption{The mean $\cal R$ for newly defined fast, medium, and slow bars with different bar length measurements \label{Table4.1.2}}
 \tablehead{
 \colhead{classification} &
 \colhead{$R_{\epsilon}$} &
 \colhead{$R_{\rm A2}$} &
 \colhead{$R_{\rm Qb}$} \\ 
}
\startdata
fast bar    & 0.54 ($\pm$0.16) & 0.71 ($\pm$0.27) & 0.61 ($\pm$0.15) \\
medium bar & 1.42 ($\pm$1.30) & 1.60 ($\pm$1.10) & 1.43 ($\pm$0.52) \\ 
slow bar  & 2.16 ($\pm$0.83) & 3.50 ($\pm$1.16) & 2.81 ($\pm$0.59) \\ 
\enddata
\end{deluxetable}

\subsection{Bar Properties and Host Galaxy} \label{chap4.2}
\begin{figure*}[htb]
\includegraphics[bb = 00 545 400 790,  width = \linewidth, clip =]{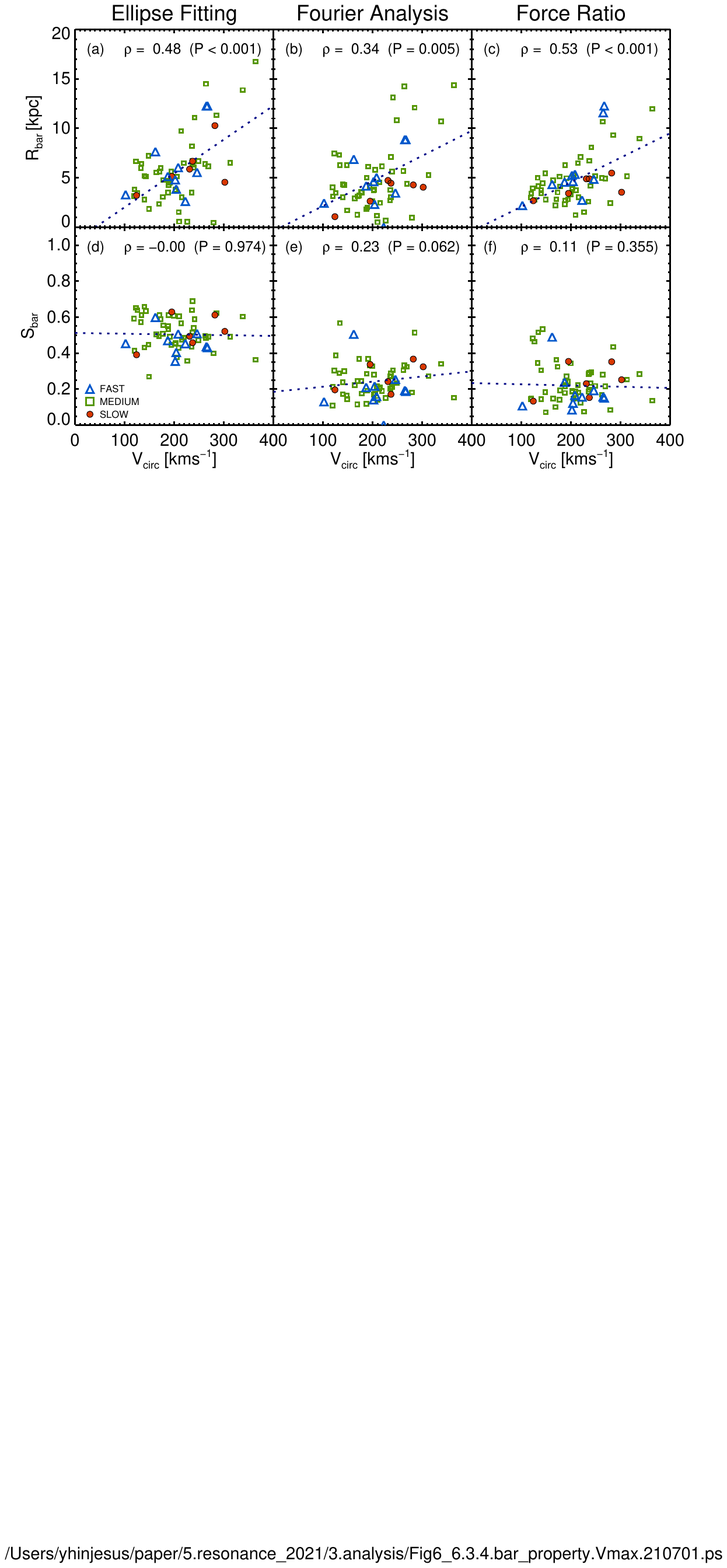}
\caption{Dependence of bar properties on the disk circular velocity $V_{\rm circ}$ as an indicator of their host property. The bar length $R_{\rm bar}$ (top row) and strength $S_{\rm bar}$ (bottom row) are measured by ellipse fitting (left), Fourier analysis (middle), and force ratio (right). The correlation is represented by Spearman's ($\rho$) with the significance (P) in the top right at each panel. The blue dotted lines show the linear fit between two parameters. The blue triangles, green squares, and solid red circles denote the newly defined fast, medium, and slow bars classified from the definition of \citet{1994Byrd}.
}\label{Fig6}
\end{figure*}

Figure \ref{Fig6} shows the bar length and strength as a function of the disk circular velocity $V_{\rm circ}$, which is a parameter tightly correlated with the galaxy luminosity through the Tully-Fisher relation \citep[TF relation,][]{1977Tully}. We obtain the disk circular velocity derived from spectroscopy in the literature \citep{2019Guo, 2019Cuomo, 2020Garma-Oehmichen}. We calculate the bar length $R_{\rm bar}$ (top row) and strength $S_{\rm bar}$ (bottom row) using the ellipse fitting ($R_{\epsilon}$ and $\epsilon_{\rm max}$ in the left), Fourier analysis ($R_{\rm A2}$ and $A_{\rm 2}$ in the middle), and force ratio ($R_{\rm Qb}$ and $Q_{\rm b}$ in the right). All the calculations are conducted following \citet{2020Lee} except the definition of $A_{\rm 2}$, which is 
\begin{equation}
A_{\rm 2} = \rm max\left(\frac{\sqrt{a_2^2+b_2^2}}{a_0}\right)    
\end{equation}
where $a_0$, $a_2$, and $b_2$ are the Fourier coefficients; this is to compare the results with more studies \citep{2013Athanassoula, 2019Seo}. We note another indicator of bar strength, Max($\Delta \mu$), the maxiumum difference between luminosity profiles along
and perpendicular to the bar axis, which correlates well with $A_2$ \citep{2017Buta, 2021Kim}.

The top row shows that the bar length depends on the circular velocity of its host galaxy. The dependence becomes strongest when we measure the bar length by the force ratio $r_{\rm Qb}$ (Figure \ref{Fig6}(c)). Previous studies reported that the bar length depends on several galaxy properties including galaxy luminosity \citep{1979Kormendy, 2020Cuomo}, effective radius, disk scale length \citep{1987Ann, 2019Erwin}, and stellar mass \citep{2016Diaz, 2019Erwin}. In particular, \citet{2019Erwin} showed that the bar length is a strong function of galaxy size in terms of effective radius $R_{\rm e}$ or disk scale length $h_{\rm r}$. He also showed an additional dependence of the bar length on galaxy mass for massive galaxies.

On the other hand, in the bottom row, we find hardly any dependence of the bar strength on the circular velocity of the host galaxy. In previous studies, the maximum ellipticity $\epsilon_{\rm max}$ appears constant across the Hubble type sequence \citep{2007Marinova, 2007Menendez, 2016Diaz, 2020Lee}. \citet{2016Diaz} and \citet{2020Lee} reported the opposite tendencies of $A_2$ and $Q_{\rm b}$ at both ends of the Hubble sequence. It is because the measurements of $A_2$ and $Q_{\rm b}$ are influenced in opposite directions by a large bulge \citep{2020Lee}. However, we find similar distributions of $A_2$ and $Q_{\rm b}$ on the circular velocity for our sample constrained by $T \le 5$ due to the limit for applying to the TW method (Figure \ref{Fig6}(e) and (f)). \citet{2020Cuomo} also reported no correlation between the bar strength and the galaxy luminosity estimated with $A_2$ for their sample galaxies analyzed by the TW method. However, it is interesting that strong bars measured with $A_2$ and $Q_{\rm b}$ are prominent in galaxies with lower velocity, $V_{\rm circ} \sim 150~ \rm km~s^{-1}$ (Figure \ref{Fig6}(e) and (f)). It seems different from the distribution of long bars, which are hosted by galaxies with higher velocity, $V_{\rm circ} > 250~\rm km~s^{-1}$ (top row).

\begin{figure*}[htb]
\includegraphics[bb = 25 640 485 790,  width = \linewidth, clip =]{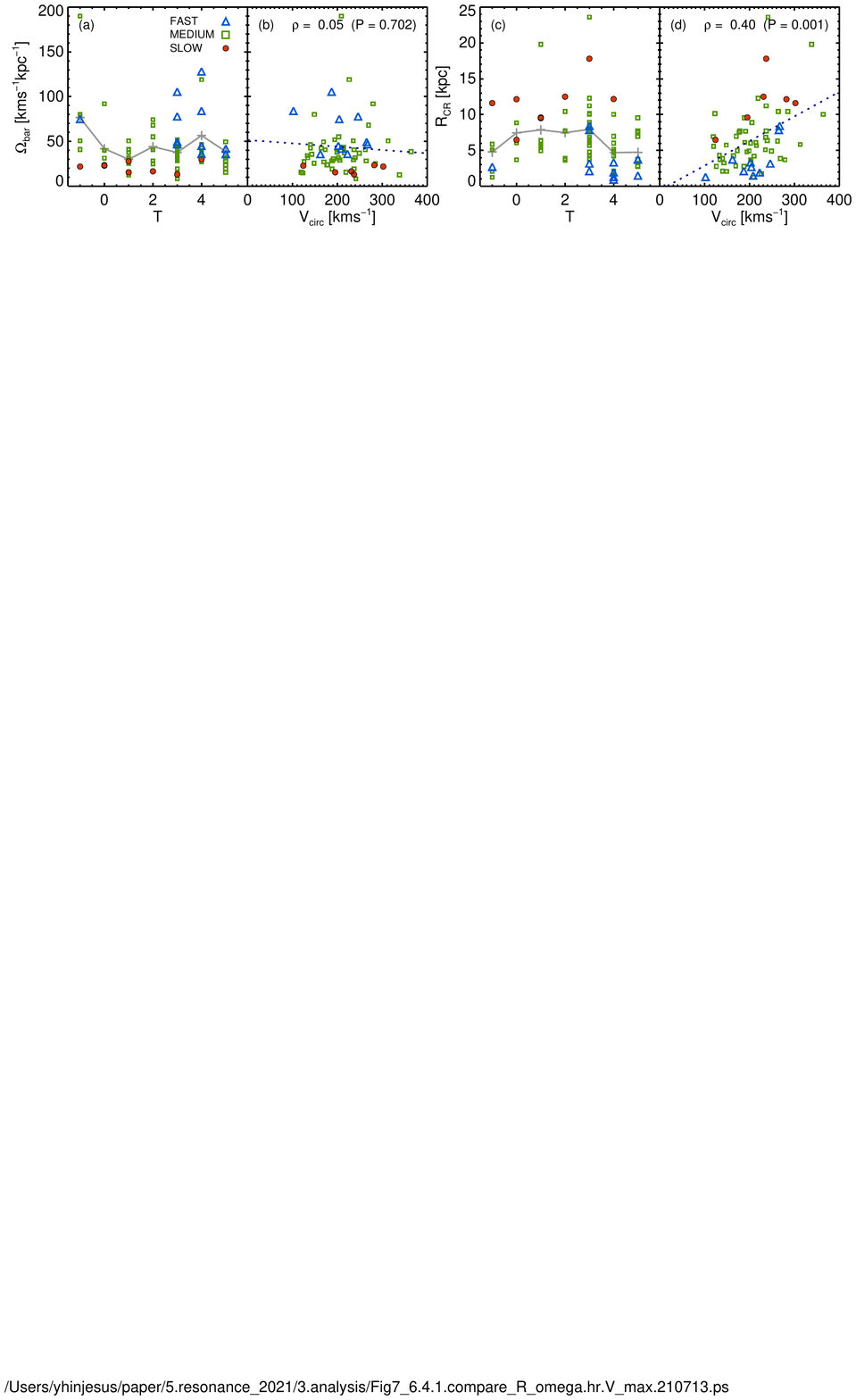}
\caption{The bar pattern speed $\Omega_{\rm bar}$ (left) and corotation radius $R_{\rm CR}$ (right) as a function of the Hubble type $T$ and the disk circular velocity $V_{\rm circ}$. We present the mean values for the Hubble type (gray solid lines) and linear fits for the correlations with the disk circular velocity (blue dotted line) together with the Spearman's coefficient ($\rho$) and its significance (P). The newly defined fast, medium, and slow bars are distinguished by blue triangles, green squares, and solid red circles.
}\label{Fig7}
\end{figure*}

\begin{figure*}[htb]
\includegraphics[bb = 30 645 460 770,  width = \linewidth, clip =]{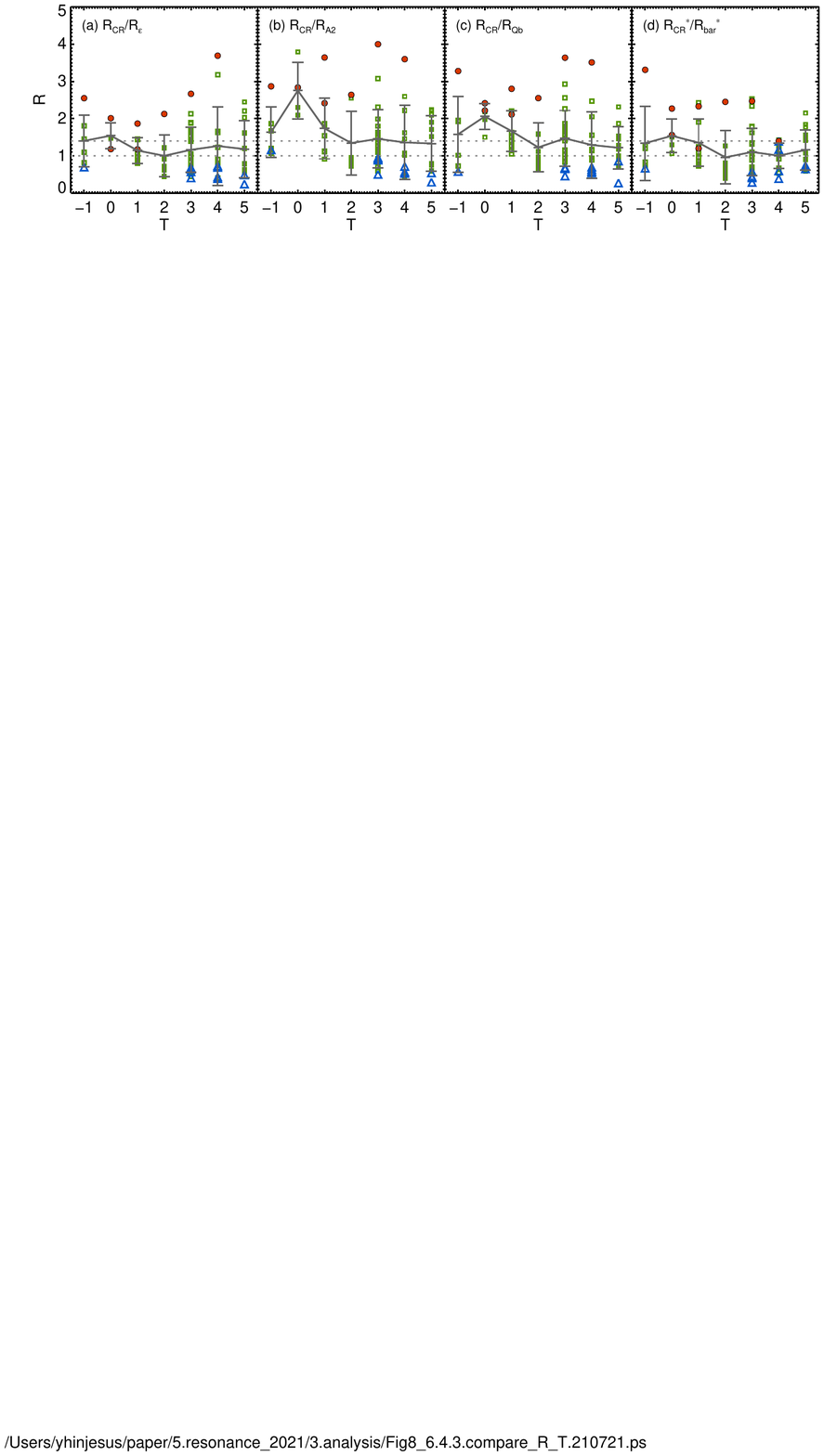}
\caption{Distribution of the ratio $\cal R$ as a function of the Hubble type for the three measures of bar length: (a) ellipse fitting, (b) Fourier analysis, and (c) force ratio. In (d), we present the ratio $\cal R$ derived from the corotation radius ${R_{\rm CR}}^\ast$ and the bar length ${R_{\rm bar}}^\ast$ in the literature. We display the mean value of $\cal R$ with error bars at each bin. The blue triangle, green square, and solid red circle represent newly defined fast, medium, and slow bars. The gray dotted horizontal lines indicate $\cal R \rm =1$ and $\cal R \rm =1.4$ distinguishing ultrafast, fast, and slow regimes designated by $\cal R$.}
\label{Fig8}
\end{figure*}

Figure \ref{Fig7} displays other important properties of bars, bar pattern speed $\Omega_{\rm bar}$ and corotation radius $R_{\rm CR}$, as a function of Hubble type $T$ and disk circular velocity $V_{\rm circ}$. We present the mean values for the Hubble type (gray solid lines) and linear fits with disk circular velocity (blue dotted line). We find that the bar pattern speed has no significant dependence either on the Hubble type or on the disk circular velocity (Figure \ref{Fig7}(a) and (b)), even though there is an S0 galaxy that has an exceptionally high pattern speed. On the other hand, the corotation radius shows a weak correlation with the disk circular velocity (Figure \ref{Fig7}(d)). When it comes to the Hubble type, Figure \ref{Fig7}(c) shows that the earlier-type spirals ($0 \le T \le 3$) have a larger corotation radius than the later-type spirals ($4 \le T \le 5$).

When we investigate this in terms of the new classification of fast (blue triangle), medium (green square), and slow (solid red circle) bars, Figure \ref{Fig6} does not show any significant difference in the correlation between bar properties (the length and the strength) and the disk circular velocity. Galaxies with different types are blended in a wide range of disk velocities. However, in Figure \ref{Fig7}, fast bars are distinguished by having a higher bar pattern speed and a smaller corotation radius contrary to slow bars. In particular, slow bars have the largest corotation radius for a given Hubble type bin or a specific velocity of host galaxies (Figure \ref{Fig7}(c) and (d)). 

The (a) and (c) panels in Figure \ref{Fig7} show that fast bars are concentrated in the later-type spirals ($T \ge 3$), whereas slow bars are distributed throughout the Hubble sequence. \citet{2008Rautiainen} reported the opposite results: earlier-type spirals have only fast bars, whereas later-type spirals host both fast and slow bars, though the definition of fast and slow bars are not the same. In terms of $\cal R$, they showed that later-type spirals have a larger value of $\cal R$. In Figure \ref{Fig8}, we also compare $\cal R$ with the Hubble type using different bar length measurements, ellipse fitting (a), Fourier analysis (b), and force ratio (c). Figure \ref{Fig8}(d) shows the galaxies based on the measurements of $R_{\rm CR}^\ast/R_{\rm bar}^\ast$ from the literature. In our measurements, the mean $\cal R$ appears slightly larger in earlier-type spirals ($T \le 1$), even though later-type spirals have a wider range of $\cal R$. On the other hand, other studies have reported no correlation between $\cal R$ and the Hubble type \citep{2015Aguerri, 2020Garma-Oehmichen, 2020Cuomo}. It seems to require a much larger sample size to better understand the relation between the ratio $\cal R$ and the Hubble type.
\subsection{Relations Between Bar Properties} \label{chap4.3}

\begin{figure*}[htb]
\includegraphics[bb = 20 445 490 800,  width = \linewidth, clip =]{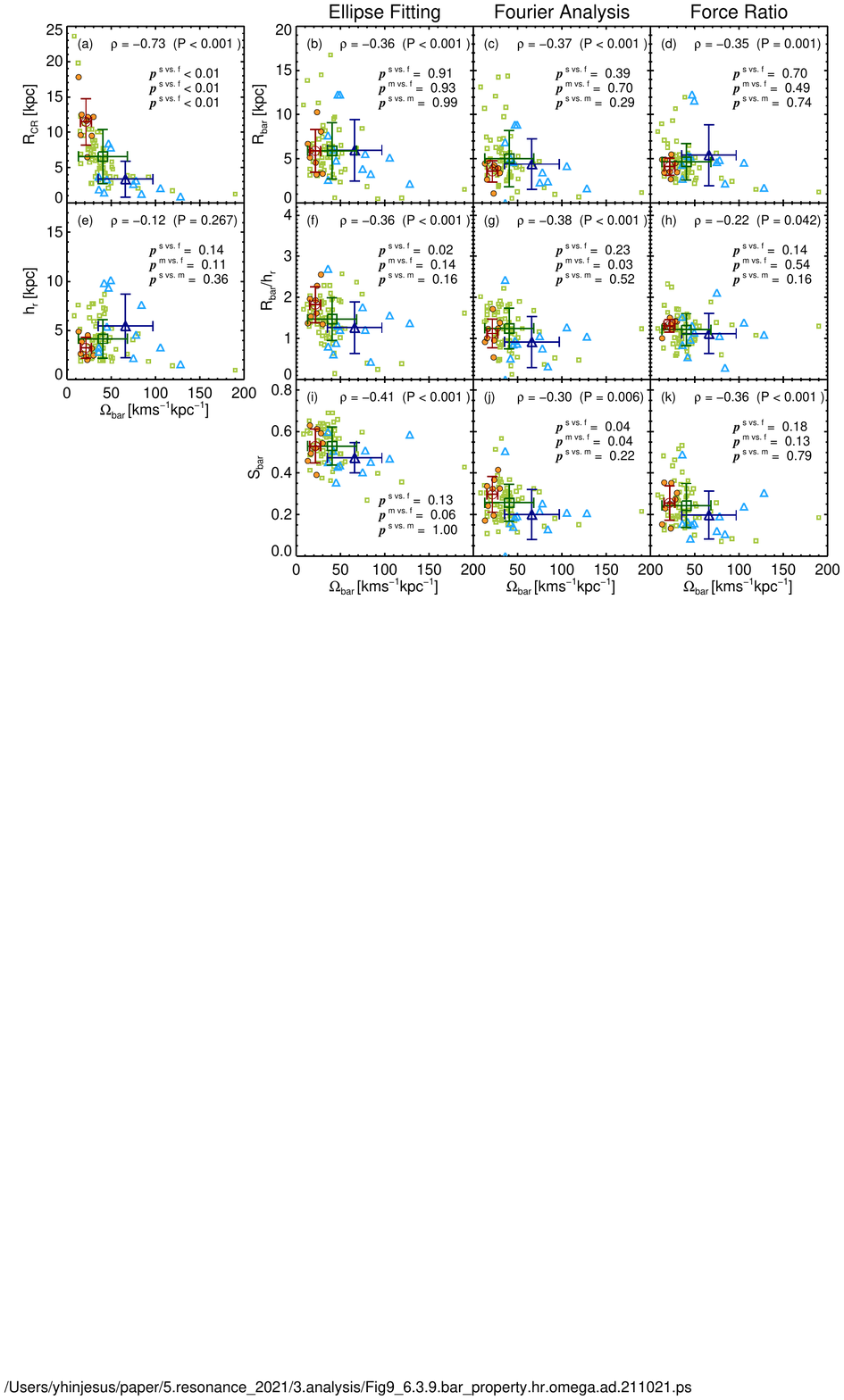}
\caption{Relations between the bar pattern speed $\Omega_{\rm bar}$ and other properties of bars, corotation radius $R_{\rm CR}$ (a), bar length  $R_{\rm bar}$ (top and middle rows), and strength $S_{\rm bar}$ (bottom row). The bar length and strength are measured by ellipse fitting (left), Fourier analysis (middle), and force ratio (right). The bar length is represented in absolute scale (top row) and in a scale normalized by the disk scale length $h_r$ (middle row). The relation between the bar pattern speed and the disk scale length $h_{\rm r}$ is shown in panel (e). The Spearman's $(\rho)$ correlation coefficient is presented with significance (P) in each panel. The newly defined fast, medium, and slow bars are denoted by the blue triangle, green square, and solid red circle. The mean and standard deviation $(\sigma)$ for each class are represented by the same color. The probability ({\it p}) from the Anderson-Darling test for the property index in the ordinate on the newly defined fast, medium, and slow bars is displayed in each panel. The superscripts, s vs. f, m vs. f, and s vs. m, stand for the two groups, slow vs. fast, medium vs. fast, and slow vs. medium bars, respectively. }\label{Fig9}
\end{figure*}

Figure \ref{Fig9} displays relations between the bar pattern speed $\Omega_{\rm bar}$ and other properties of bars including corotation radius $R_{\rm CR}$, bar length $R_{\rm bar}$, and strength $S_{\rm bar}$. We present the bar lengths in an absolute scale (top row) and in a scale normalized by the disk scale length $h_{\rm r}$ (middle row). The relation between the bar pattern speed and the disk scale length is displayed at the left most panel in the middle row. 

First, we find that the bar pattern speed $\Omega_{\rm bar}$ is anti-correlated with other properties of bars: as the bar pattern speed decreases, the values of other parameters increase. Figure \ref{Fig9}(a) shows that the pattern speed is anti-correlated with the corotation radius (confirmed by the $\rho = -0.73$). It is expected because the disk angular velocity decreases in proportion to $r^{-3/2}$ so that a large corotation radius for low pattern speed. On the other hand, the pattern speed has weak anti-correlations with the bar length and the strength with $\rho \sim -0.35$. It appears to support the concept of bar growth in terms of length and strength through the slowdown of a bar by losing its angular momentum from a dark halo \citep{2000Debattista, 2003Athanassoula, 2019Seo}.

However, if the relation between the bar pattern speed and the frequency curves gives a hint for the evolutionary stage of barred galaxies, we can investigate the relations between the bar pattern speed and other bar properties with another view. In Figure \ref{Fig9}, we present the mean values with error bars for fast (blue triangle), medium (green square), and slow bars (solid red circle). The panel (a) shows that slow bars have lower pattern speed and larger corotation radius than fast and medium bars as expected. However, we cannot find any increase in the bar length from fast bars to medium or slow bars (top row). In the case of $R_{\rm Qb}$, it even decreases from fast bars to slow bars (Figure \ref{Fig9}(d)). When we investigate the normalized bar length, it shows a tendency of larger bar lengths for slowly rotating bars (middle row). However, it is caused by the decrease of disk scale length from fast to slow bars, as shown in panel (e). We note that the disk scale length also could be changed between fast and slow bars. For the bar strength, we find a weak tendency of increasing bar strength from fast to slow bars, even though the increases are within error bars (bottom row). 

To examine the difference among new subclasses of fast, medium and slow bars, we perform the Anderson-Darling (A-D) test for each combination of subclasses. We list the relevant {\it p}-value in each panel, which indicates the probability that the two samples  are drawn from the same parent distribution. Firstly, {\it p}-values of the A-D test for the pattern speed distributions ($\Omega_{\rm bar}$) between fast and medium bars and between medium and slow bars are 0.009 and 0.002, respectively. This means that the newly defined subclasses of bars is relevant to the differentiation of the pattern speed. Secondly, the probability from the A-D test for each property index, $R_{\rm CR}$, $h_{\rm r}$, $R_{\rm bar}$, $R_{\rm bar}/h_{\rm r}$, and $S_{\rm bar}$, is noted as {\it p} in each panel, which shows that the three subclasses show different distributions in the corotation radius, but do not show differences in other properties. We will discuss these results on the bar evolution in Section \ref{chap5.3}.

\section{Discussion}\label{chap5}
\subsection{Bar Length and Resonance} \label{chap5.1}
\begin{figure*}[htb]
\includegraphics[bb = 20 640 425 790,  width = \linewidth, clip =]{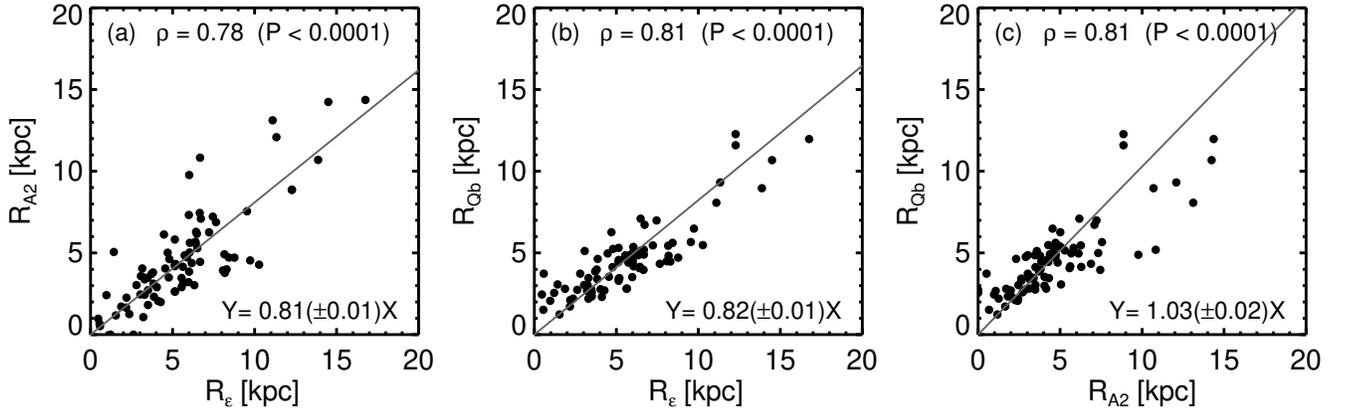}
\caption{Comparison between the bar lengths measured by different methods, ellipse fitting ($R_{\epsilon}$), Fourier analysis ($R_{\rm A2}$), and force ratio ($R_{\rm Qb}$). We present the Spearman's ($\rho$) correlation coefficient with the significance (P). The solid line denotes the linear fit between bar lengths from different methods.}\label{Fig10}
\end{figure*}

In this work, we have used three measures of bar length defined by the radius, $R_{\epsilon}$, $R_{\rm A2}$, and $R_{\rm Qb}$, where the ellipticity ($\epsilon$), Fourier amplitude ($A_{\rm 2}$), and force ratio ($Q_{\rm b}$) reach their maxima in the radial profiles. Figure \ref{Fig10} shows correlations between bar length measurements. All the three measures of bar length are strongly correlated with each other; in particular, $R_{\rm A_2}$ and $R_{\rm Q_b}$ are similar to each other, resulting in the slope of one for the correlation between the two (Figure \ref{Fig10}(c)). However, they are measured to be shorter than $R_{\epsilon}$ by 20\% (Figure \ref{Fig3}(a) and (b)). \citet{2016Diaz} reported that $R_{\epsilon}$ is the best indicator of the visually estimated bar length.

Figure \ref{Fig4} shows four example galaxies with three bar length measurements overlaid on their images. We present the positions of the tip of the bar measured from ellipse fitting (red), Fourier analysis (green), and force ratio (blue) on the images, in the same manner as \citet{2021Cuomo}. When we investigate the bar length measurements on images one by one, all three measurements are compatible for galaxies with simple structures such as shown in Figure \ref{Fig4}(a). However, when a galaxy hosts a pseudo ring or ring around a bar, $R_{\epsilon}$ tends to be located on the ring (Figure \ref{Fig4}(b)-(d)): the ellipticity gradually increases up to the ring radius. It could make the bar length overestimated \citep{2021Cuomo}. On the contrary, in Figure \ref{Fig4}(d), $R_{\rm A_2}$ and $R_{\rm Q_b}$ are located at a very inner region despite a long bar that is visible in the image. This may mean that the maximum radius of $A_2$ and $Q_b$ cannot reflect the bar length in the case of a strong bar, in particular. We also note that $R_{\rm A_2}$ could be easily affected by spiral arms  \citep{2002Laurikainen, 2004aLaurikainen}. \citet{2003Buta} introduced an extrapolation method to separate a bar from spiral arms on the Fourier amplitude profile. In this work, we set a radius limit in finding the maximum $A_{\rm 2}$ to avoid the contamination from spiral arms through a visual inspection.

\begin{figure*}[htb]
\includegraphics[bb = 0 335 390 790,  width = \linewidth, clip =]{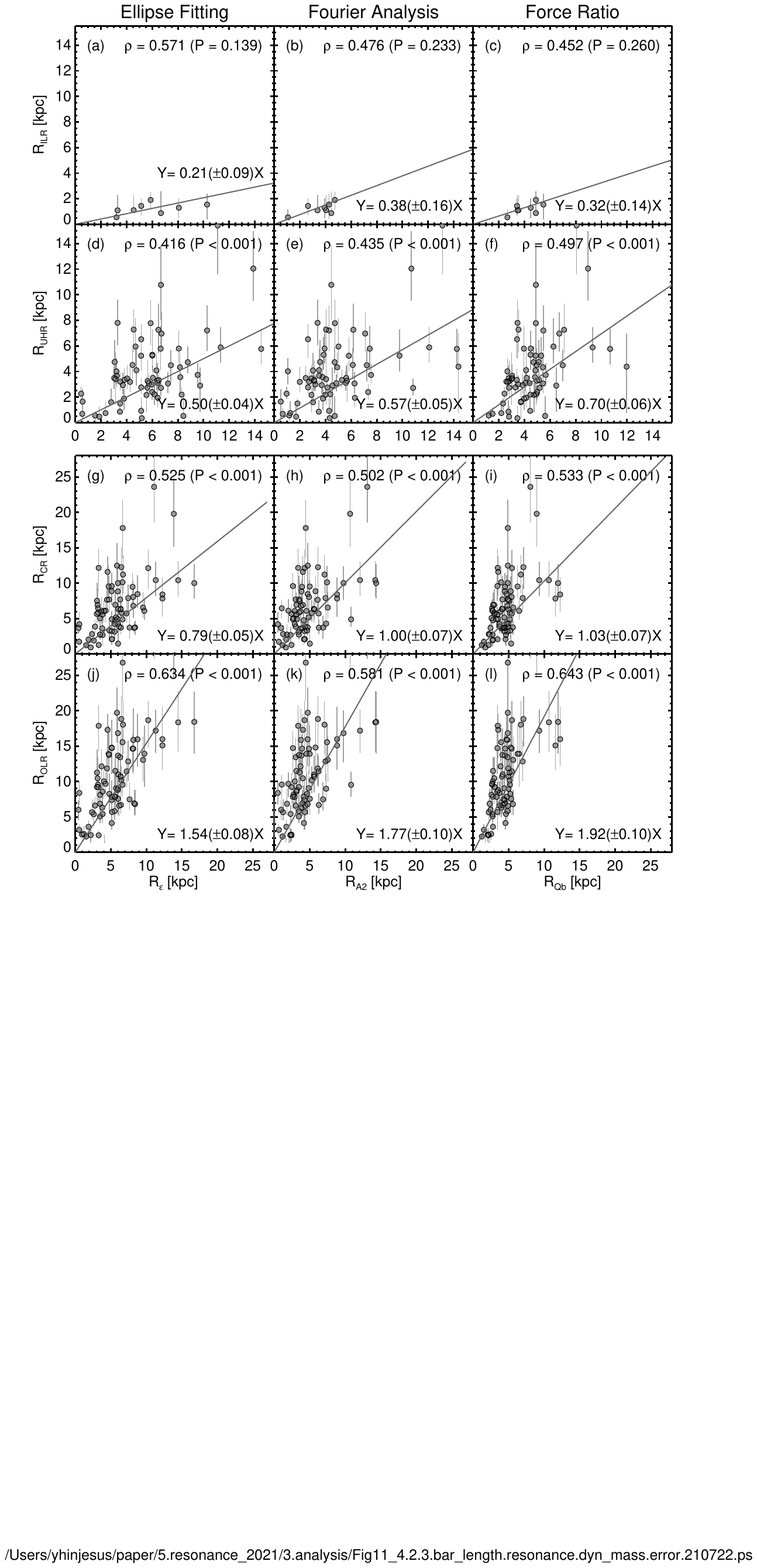}
\caption{Correlation between the resonance radii and the bar lengths from different methods. The resonance radii of ILR, UHR, CR, and OLR are shown from top to bottom. The different bar length measurements are displayed from left (ellipse fitting) to middle (Fourier analysis) and right (force ratio). The Spearman's ($\rho$) coefficients are presented with the significance (P) at the top right. The linear fit between two parameters, shown at the bottom right, is displayed by the solid line in each panel.}\label{Fig11}
\end{figure*}

From N-body simulations, \citet{2006Michel} showed that the bar lengths measured by different methods are correlated with different resonances. They investigated various radii available to define the bar length where a maximum, a minimum, or a transition between a bar and a disk appears on the ellipticity or Fourier amplitude profile. They showed that the bar length defined as the radius of a minimum ellipticity corresponds to the corotation radius (CR), while the bar length defined by the transition between a bar and a disk locates close to the Ultraharmonic Resonance (UHR). They reported that the length measured by force ratio, $R_{\rm Qb}$, is located in the circumnuclear region, and the length measured by maximum ellipticity does not show any correlation with dynamical resonances. As a result, they considered that the bar length measured by the maximum ellipticity, $R_{\epsilon}$, is not a proper estimator of the bar length. 

Therefore, we compare the correlations between the bar length estimates and the dynamical resonance locations for our whole sample in Figure \ref{Fig11}. $R_{\rm ILR}$, $R_{\rm UHR}$, $R_{\rm CR}$, and $R_{\rm OLR}$ are measured where the pattern speed of a bar $\Omega_{\rm bar}$ intersects the frequency curves, $\Omega-\kappa/2$, $\Omega-\kappa/4$, $\Omega$, and $\Omega+\kappa/2$, in sequence. The plot shows $R_{\rm ILR}$, $R_{\rm UHR}$, $R_{\rm CR}$, and $R_{\rm OLR}$ from top to bottom and the bar length measurements by $
R_{\epsilon}$, $R_{\rm A2}$, and $R_{\rm Qb}$ from left to right. We display Spearman's correlation coefficient ($\rho$) with the significance (P) at the top right and the best-fit relation between the resonance radii and the bar length with the solid line.

The plot shows that all the dynamical resonances are strongly correlated with the bar lengths regardless of the method to measure the bar length. The correlations between the bar length and the resonance locations seem to be tighter for the resonances such as OLR or CR. The bar length by $R_{\rm A2}$ and $R_{\rm Qb}$ locates near CR (Figure \ref{Fig11}(h) and (i)), but the scatter is very large. Compared to the results in \citet{2006Michel}, our resonance radii tend to be located inward, because both the minimum ellipticity and the transition between a bar and a disk occur after the maximum ellipticity. In conclusion, we hardly find the different links with specific dynamical resonances for different bar length measurements. The simulations may not be sufficient to compare with observations because few are available. We need more simulation models with various properties for detailed comparison.

\subsection{Evolution of Barred Galaxies in terms of $\cal R$} \label{chap5.2} 

In Section \ref{chap4.1.1}, we introduced that most barred galaxies with measured bar pattern speeds belong to fast bars in terms of $\cal R$ \citep{2015Aguerri, 2019Cuomo, 2020Garma-Oehmichen}. The observed galaxies with lower $\cal R$ or with a small number of slow bars have led to concerns about less-concentrated dark matter halo or inefficient angular momentum exchange between a bar and a dark halo \citep{2012Perez, 2015Aguerri}. First, we suggest that the different bar length measurements can more or less explain the discrepancy of $\cal R$ between observations and simulations. Because observations usually used the ellipse fitting method to measure the bar length \citep{2015Aguerri, 2019Guo, 2019Cuomo, 2020Garma-Oehmichen}, while simulations obtained the bar length from Fourier anlysis \citep{1996aLindblad, 2000Debattista, 2013Athanassoula, 2019Seo}. In our measurements, $R_{\rm A2}$ from Fourier analysis is shorter than $R_{\epsilon}$ by 20\%, which can explain a larger $\cal R$ in simulations.

Secondly, we argue that a certain criterion of $\cal R \rm =1.4$ may not be appropriate to classify barred galaxies into fast and slow bars or to constrain dark halo density. The simulation that suggested $\cal R \rm \le 1.4$ only for a dark halo with low density did not consider the existence of gas in their simulations \citep{2000Debattista}. However, \citet{2014Athanassoula} showed that barred galaxies even in a dense halo evolve within $\cal R \rm =1.4$ when they have enough gas initially. Moreover, the shape or spin of a halo influences the evolution of barred galaxies: triaxial or fast spinning haloes drive lower $\cal R$ \citep{2014Athanassoula, 2014Long, 2018Collier}.  

Nevertheless, observations show large differences from the simulation results, including the EAGLE and the Illustris TNG  projects. \citet{2021Roshan} obtained $\cal R \rm \sim 2.5$, in average, for simulated galaxies of EAGLE and Illustris by measuring the bar pattern speed and the bar length through the TW method and $R_{\rm A2}$. The simulated galaxies have a much longer corotation radius over $10~\rm kpc$ and a smaller bar length ($<R_{\rm A2}> = 3.1~{\rm kpc}$) than those in observations. \citet{2017Algorry} also reported that strong bars in EAGLE simulations have corotation radii lager than 10 times the bar length. However, we cannot find such highly evolved barred galaxies in observations. The IllustrisTNG simulations also yield a slower bar pattern speed $<\Omega_{\rm bar}> = 25.2~{\rm km~s^{-1}~kpc^{-1}}$ \citep{2021Roshan} than those of our sample galaxies, which have $<\Omega_{\rm bar}> = 44.1 \pm 29.1 \rm ~km~s^{-1}~kpc^{-1}$.

\subsection{Little Secular Evolution of Barred Galaxies} \label{chap5.3}

In Section \ref{chap4.2}, we examined the dependence of the bar length on the disk circular velocity of host galaxy, and found that rapidly rotating disks host long bars (Figure \ref{Fig6}). The Tully-Fisher relation \citep{1977Tully} dictates that rapidly rotating disks are luminous and massive. The observation of longer bars in brighter, massive, and larger galaxies in the local universe \citep{1979Kormendy, 1987Ann, 2019Erwin, 2020Cuomo} could be an outcome of various processes. Long bars could inherit their size from their host galaxies; larger and massive galaxies could make their bars evolve longer effectively. Host galaxies and bars could evolve together by mutual interactions. In any case, we need to consider the disk velocity when we investigate the evolution of barred galaxies. 

In Figure \ref{Fig4}, we showed examples of a nonbarred galaxy along with those of fast, medium, and slow bars. We selected them by fixing their velocity $V_{\rm circ} = 190 \pm 10~\rm km~s^{-1}$ except nonbarred galaxies. In our sample, nonbarred galaxies without a CR are mainly slowly rotating galaxies with $V_{\rm circ} < 150~\rm km~s^{-1}$. The circular velocity of UGC 3944 is $148~\rm km~s^{-1}$ (Figure \ref{Fig4}(a)). In the bottom row, we present the radial profile of the force ratio for each galaxy. \citet{2020Lee} introduced a way to analyze a force ratio map defined as the transverse-to-radial force ratio by investigating the radial and azimuthal profiles. From a comparison with simulations, they suggested an evolution process of barred galaxies on the radial profile of force ratio. Galaxies grow from a plateau (type-P) to a maximum peak (type-M) on the radial profile with increasing force ratio $Q_{\rm b}$ \citep[see their Figure 19]{2020Lee}. The galaxies in Figure \ref{Fig4} seem to follow the evolution process suggested in \citet{2020Lee}: the radial profiles show a plateau for a nonbarred galaxy (Figure \ref{Fig4}(i)), whereas fast, medium, and slow bars have a maximum peak on the radial profile (Figure \ref{Fig4}(g)-(l)). They show increasing force ratios $Q_{\rm b}$ from a fast bar to a slow bar. On the other hand, we hardly find the increase in the bar length from a fast bar to a slow bar. In particular, the maximum radii of $A_{\rm 2}$ and $Q_{\rm b}$ seem to be located inward compared to the bar end in the slow bar.

In Figure \ref{Fig9}, the anti-correlation between the bar pattern speed and the bar length seems to show the growth of the bar length as the bar pattern speed decreases through an exchange of angular momentum between a bar and a dark halo \citep{2002Athanassoula, 2003Athanassoula}. However, we are concerned that all of the longer bars with $R_{\rm bar} > 10$ kpc are found only in rapidly rotating systems, namely massive galaxies. When we investigate them by classifying into fast, medium, and slow bars, we can not find any difference in the bar length between fast and slow bars. Although the normalized bar length shows a trend to increase from a fast bar to a slow bar, it is caused by the decrease in the disk scale length. Therefore, long bars in massive galaxies seem to inherit the size from their host galaxy where they form. The bar instability in a massive disk galaxy may yield a large massive bar from the beginning of bar formation. When we normalize the bar length by indicators of the galaxy size, we need to be careful that the disk scale length could be changed as well during the bar evolution. When it comes to the bar strength, we can find a hint of increase by evolution, but the amount of increase is not large.

In conclusion, we do not find the increase of bar length and strength for the bar evolution predicted by numerical simulations \citep{2000Debattista, 2003Athanassoula, 2019Seo}. It is in line with previous observations that most galaxies stay in the phase of fast bars in terms of $\cal R$ \citep{2015Aguerri, 2019Guo, 2019Cuomo}. The recent observational study of \citet{2021Kim} supports little secular evolution of barred galaxies as well. They investigated the evolution of bar length and strength at $0.2 < z \le 0.835$ from HST/COSMOS data. They showed that the absolute and normalized bar lengths have rarely been changed over the last 7 Gyr. They found only a slight increase of the bar strength over cosmic time. Therefore, they discussed the cases of simulation models that bars could experience very little secular evolution, including gaseous disk, triaxial halo \citep{2013Athanassoula}, or increasing dark halo spin \citep{2014Long}. \citet{2015Okamoto} also showed that bars do not always grow by evolution in self-consistent hydrodynamical simulation for two Milky Way-mass galaxies in cosmological context.

\section{Summary} \label{chap6}
We have derived the stellar frequency of the circular orbit $\Omega$ and the epicyclic precession frequencies, $\Omega \pm \kappa/2$ and $\Omega-\kappa/4$ for barred galaxies from photometry. We constructed mass maps using the dynamical mass-to-light ratio from a surface brightness distribution and a galaxy color. The gravitational potential is calculated by solving the Poisson equation for the mass map. We determined the resonance locations, ILR, UHR, CR, and OLR, by directly putting the bar pattern speed on the frequency curves. We utilized the bar pattern speed measured with the TW method from Integral Field Spectroscopy (IFS) data in the literature. 

Our main results are summarized as follows,
\begin{enumerate}
\item We show that the ratio $\cal R \rm = R_{\rm CR}/R_{\rm bar}$ depends on the method of bar length measurements. The bar length from the Fourier analysis and the force ratio are measured to be smaller than that from the ellipse fitting by 20\%. It explains, at least partly, the larger $\cal R$ values in simulations that usually used the Fourier analysis to measure the bar length. 

\item We take a different approach to classify barred galaxies into fast, medium, and slow bars by putting the bar pattern speed on the frequency curves. It reflects an evolutionary process that lower pattern speeds by losing angular momentum intersect with more frequency curves. We found 11 fast, 59 medium, and 8 slow bars in this way even though we might have missed some of slow bars due to the resolution limit. Five galaxies have no corotation radius because of high bar pattern speed not intersecting the angular velocity curve.

\item We find that the bar length and corotation radius depend on the disk circular velocity of its host galaxy, while the bar strength and the pattern speed are independent of the disk circular velocity. Long bars are found at galaxies with higher velocity, $V_{\rm circ} > 250~\rm km~s^{-1}$. However, strong bars are prominent in galaxies with lower velocity $V_{\rm circ} \sim 150~ \rm km~s^{-1}$.

\item The bar pattern speed is anti-correlated with other properties of bars: as the bar pattern speed decreases, the corotation radius, the bar length, and the strength increase. However, if we divide the galaxies into fast, medium, and slow bars, there is no increase in the bar length. We only find a hint of the increase in the strength. The bars in galaxies seem to experience little evolution in terms of bar length and strength. 

\end{enumerate}

We thank the reviewer for detailed and insightful comments on the manuscript, which greatly improved the paper. MGP acknowledge support from the Basic Science Research Program through the National Research Foundation of Korea (NRF) funded by the Ministry of Education (No. 2019R1I1A3A02062242). HSH was supported by the New Faculty Startup Fund from Seoul National University. TK was supported by the Basic Science Research Program through the National Research Foundation of Korea (NRF) funded by the Ministry of Education (NRF-2019R1A6A3A01092024).

\bibliography{resonance_ref}
\appendix

\setcounter{table}{0}
\renewcommand{\thetable}{A\arabic{table}}

\begin{longrotatetable}
\begin{deluxetable*}{l|rrrrrrr|rrrr|l}
\tablecaption{Main parameters and resonance locations of our sample \label{Table1}}
\tabletypesize{\scriptsize}
\tablehead{
\colhead{Galaxy} & \colhead{RA} & \colhead{Dec.} & \colhead{distance} & \colhead{Morph.} & \colhead{incl.} & \colhead{PA} & \colhead{$\Omega_{\rm bar}$} & \colhead{$R_{\rm ILR}$} & \colhead{$R_{\rm UHR}$} & \colhead{$R_{\rm CR}$} & \colhead{$R_{\rm OLR}$} & \colhead{ref.} \\
\colhead{} & \colhead{[$^\circ$]} & \colhead{[$^\circ$]} & \colhead{[Mpc]} & \colhead{} & \colhead{[degree]} & \colhead{[degree]} & \colhead{[$\rm kms^{-1}kpc^{-1}$]} & \colhead{[kpc]} & \colhead{[kpc]} & \colhead{[kpc]} & \colhead{[kpc]} & 
\colhead{} \\
\colhead{(1)} & \colhead{(2)} & \colhead{(3)} & \colhead{(4)} & \colhead{(5)} & \colhead{(6)} & \colhead{(7)} & \colhead{(8)} & \colhead{(9)} & \colhead{(10)} & \colhead{(11)} & \colhead{(12)} & \colhead{(13)}}

\startdata
    NGC 36 & 2.8429 & 6.3894 &  83.65 &   SBb &    57.20 &    23.40 & $ 32.55_{ -9.12}^{+  9.12}$ & $  - $ & $  4.33_{ -0.86}^{+  1.19}$ & $  8.06_{ -2.03}^{+  2.31}$ & $ 14.    65_{ -3.58}^{+  3.85}$ & A \\
    NGC 1645 & 71.0267 & -5.4656 &  71.68 &  SB0a &    64.50 &    84.70 & $ 31.37_{-33.38}^{+ 33.38}$ & $  - $ & $  5.31_{ -1.16}^{+  1.41}$ & $  8.85_{ -2.29}^{+  1.67}$ & $ 13.    68_{ -3.39}^{+  2.47}$ & A \\
    NGC 3300 & 159.1600 & 14.1711 &  50.45 &  SB0a &    57.20 &   172.00 & $ 40.07_{-12.67}^{+ 12.67}$ & $  - $ & $  3.45_{ -0.82}^{+  1.03}$ & $  6.09_{ -1.59}^{+  1.49}$ & $ 10.    03_{ -2.47}^{+  1.60}$ & A \\
    NGC 5205 & 202.5150 & 62.5117 &  27.52 &  SBbc &    50.00 &   170.10 & $128.17_{-32.23}^{+ 32.23}$ & $  - $ & $  - $ & $  0.93_{ -0.30}^{+  0.28}$ & $  1.    89_{ -0.47}^{+  0.70}$ & A \\
    NGC 5378 & 209.2125 & 37.7972 &  47.69 &   SBb &    37.80 &    86.50 & $ 48.44_{-26.38}^{+ 26.38}$ & $  - $ & $  2.19_{ -0.41}^{+  0.57}$ & $  3.70_{ -0.66}^{+  0.83}$ & $  6.    94_{ -1.81}^{+  1.86}$ & A \\
    NGC 5406 & 210.0837 & 38.9156 &  79.44 &   SBb &    44.90 &   111.80 & $ 42.32_{-24.15}^{+ 24.15}$ & $  - $ & $  3.73_{ -0.63}^{+  0.92}$ & $  6.57_{ -1.43}^{+  2.17}$ & $ 13.    06_{ -3.08}^{+  3.29}$ & A \\
    NGC 5947 & 232.6525 & 42.7172 &  88.14 &  SBbc &    44.60 &    72.50 & $ 38.38_{-17.08}^{+ 17.08}$ & $  - $ & $  3.09_{ -0.66}^{+  0.67}$ & $  5.18_{ -1.12}^{+  1.34}$ & $  9.    06_{ -2.09}^{+  2.54}$ & A \\
    NGC 6497 & 267.8250 & 59.4708 &  89.84 &  SBab &    60.90 &   112.00 & $ 74.39_{-12.63}^{+ 12.63}$ & $  - $ & $  0.52_{   }^{+  1.55}$ & $  3.77_{ -1.05}^{+  0.95}$ & $  6.    81_{ -1.57}^{+  2.03}$ & A \\
    NGC 6941 & 309.0979 & -4.6186 & 87.64 &   SBb &    42.30 &   127.50 & $ 30.12_{-24.48}^{+ 24.48}$ & $  - $ & $  4.73_{ -0.88}^{+  1.23}$ & $  8.47_{ -1.97}^{+  2.68}$ & $ 15.    97_{ -3.83}^{+  3.63}$ & A \\
    NGC 6945 & 309.7525 & -4.9725 &  51.81 &   SB0 &    51.30 &   126.10 & $ 41.01_{-14.33}^{+ 14.33}$ & $  - $ & $  3.67_{ -0.88}^{+  0.89}$ & $  5.86_{ -1.33}^{+  1.15}$ & $  9.    23_{ -2.34}^{+  1.68}$ & A \\
    NGC 7321 & 339.1167 & 21.6219 & 100.15 &  SBbc &    48.40 &    13.40 & $ 34.40_{ -7.83}^{+  7.83}$ & $  - $ & $  5.24_{ -1.25}^{+  2.05}$ & $ 10.03_{ -2.36}^{+  2.38}$ & $ 16.    84_{ -4.14}^{+  3.21}$ & A \\
    NGC 7563 & 348.9829 & 13.1961 &  56.12 &   SBa &    55.80 &   149.80 & $ 27.93_{-12.13}^{+ 12.13}$ & $  1.30_{ -0.83}^{+  0.73}$ & $  5.80_{ -1.14}^{+  1.39}$ & $  9.49_{ -2.38}^{+  2.00}$ & $ 14.    65_{ -3.24}^{+  2.03}$ & A \\
    NGC 7591 & 349.5679 & 6.5858 &  67.63 &  SBbc &    57.60 &   144.00 & $ 29.89_{-14.03}^{+ 14.03}$ & $  1.09_{   }^{+  1.19}$ & $  7.80_{ -1.73}^{+  1.81}$ & $ 12.17_{ -3.18}^{+  2.69}$ & $ 17.    88_{ -3.58}^{+  2.95}$ & A \\
   UGC 3253 & 79.9246 & 84.0525 & 60.27 &   SBb &    56.80 &    92.00 & $ 35.94_{-10.61}^{+ 10.61}$ & $  -$ & $  3.11_{ -0.60}^{+  0.90}$ & $  6.38_{ -1.94}^{+  1.59}$ & $ 10.    70_{ -2.28}^{+  2.13}$ & A \\
   UGC 12185 & 341.8542 & 31.3736 &   93.11 &   SBb &    64.00 &   161.00 & $ 29.46_{ -3.77}^{+  3.77}$ & $  - $ & $  4.49_{ -1.29}^{+  1.20}$ & $  7.91_{ -1.91}^{+  1.88}$ & $ 12.    85_{ -2.73}^{+  2.93}$ & A \\
     IC 1528 & 1.2724 &  -7.0934 & 49.90 & SABbc &    66.70 &    72.70 & $ 87.00_{-20.00}^{+ 20.00}$ & $  - $ & $  - $ & $  - $ & $  1.    43_{   }^{+  1.13}$ & C \\
     IC 1683 & 20.6619 & 34.4370 &  66.20 &  SABb &    54.30 &    13.00 & $ 30.30_{ -5.10}^{+  5.10}$ & $  - $ & $  4.51_{ -1.11}^{+  1.32}$ & $  7.66_{ -1.66}^{+  1.61}$ & $ 11.    85_{ -2.46}^{+  2.12}$ & C \\
     IC 5309 & 349.7985 &  8.1093 &  55.20 &  SABc &    60.00 &    26.70 & $ 91.00_{-26.00}^{+ 26.00}$ & $  - $ & $  - $ & $  - $ & $  2.    53_{ -0.93}^{+  0.77}$ & C \\
 MCG-02-02-030 & 7.5305 &    -11.1137 &  45.90 &  SABc &    36.90 &    98.50 & $ 43.40_{ -6.50}^{+  6.50}$ & $  - $ & $  1.65_{ -1.31}^{+  0.98}$ & $  4.21_{ -1.19}^{+  1.74}$ & $  8.    40_{ -1.66}^{+  1.61}$ & C \\
     NGC 551 & 21.9193 & 37.1830  &  71.10 & SABbc &    64.70 &   137.00 & $ 45.00_{-11.00}^{+ 11.00}$ & $  - $ & $  - $ & $  3.33_{ -1.30}^{+  1.28}$ & $  8.    27_{ -2.62}^{+  2.87}$ & C \\
    NGC 2449 & 116.8345 & 26.9302 &  73.30 & SABab &    69.20 &   136.40 & $ 40.70_{ -5.50}^{+  5.50}$ & $  - $ & $  4.08_{ -1.56}^{+  1.38}$ & $  7.71_{ -2.09}^{+  2.17}$ & $ 13.    44_{ -3.45}^{+  2.72}$ & C \\
    NGC 2553 & 124.3960 &   20.9032 &  71.50 & SABab &    54.60 &    67.00 & $ 68.10_{ -9.80}^{+  9.80}$ & $  - $ & $  2.22_{ -0.84}^{+  0.70}$ & $  3.87_{ -0.72}^{+  0.92}$ & $  6.    58_{ -1.50}^{+  1.52}$ & C \\
    NGC 2880 & 142.3942 &  62.4906 &  24.10 &  EAB7 &    56.70 &   144.60 & $190.00_{-28.00}^{+ 28.00}$ & $  - $ & $  0.56_{ -0.39}^{+  0.33}$ & $  1.25_{ -0.26}^{+  0.33}$ & $  2.    24_{ -0.52}^{+  0.49}$ & C \\
    NGC 3994 & 179.4036 &  32.2776 &  48.60 & SABbc &    63.00 &     6.90 & $119.00_{-27.00}^{+ 27.00}$ & $  - $ & $  0.69_{   }^{+  0.49}$ & $  1.74_{ -0.39}^{+  0.47}$ & $  3.    19_{ -0.76}^{+  0.75}$ & C \\
    NGC 6278 & 255.2097 &   23.0110 &  40.90 & SAB0 &    58.80 &   126.40 & $ 92.00_{-28.00}^{+ 28.00}$ & $  - $ & $  2.27_{ -0.41}^{+  0.54}$ & $  3.68_{ -0.76}^{+  0.76}$ & $  6.    02_{ -1.48}^{+  1.32}$ & C \\
    UGC 3944 & 114.6521 & 37.6335 &  58.30 & SABbc &    59.30 &   119.60 & $ 62.00_{-22.00}^{+ 22.00}$ & $  - $ & $  - $ & $  - $ & $  2.    45_{ -1.05}^{+  1.04}$ & C \\
    UGC 8231 & 197.1552  &  54.0745 &  37.80 &  SABd &    68.10 &    74.20 & $ 58.00_{-31.00}^{+ 31.00}$ & $  - $ & $  - $ & $  - $ & $  - $ & C \\
 manga-7495-12704 & 205.4384 & 27.0048 & 123.70 &  SBbc &    52.20 &   173.40 & $ 31.65_{ -2.95}^{+  3.80}$ & $  - $ & $  2.93_{ -0.67}^{+  0.96}$ & $  6.01_{ -1.66}^{+  2.12}$ & $ 12.    15_{ -3.16}^{+  2.49}$ & G \\
 manga-7962-12703 & 261.2173 & 28.0783 & 203.30 &  SBab &    61.20 &    32.40 & $ 28.25_{ -0.69}^{+  0.93}$ & $  - $ & $  5.77_{ -1.23}^{+  1.56}$ & $ 10.43_{ -2.35}^{+  2.71}$ & $ 18.    36_{ -4.15}^{+  4.41}$ & G \\
  manga-7990-3704 & 262.0749 & 56.7748 & 124.60 &   SB0 &    39.40 &    11.60 & $ 80.07_{-25.30}^{+ 25.56}$ & $  - $ & $  0.47_{   }^{+  0.83}$ & $  2.06_{ -0.50}^{+  0.50}$ & $  3.    62_{ -0.66}^{+  0.92}$ & G \\
   manga-7990-9101 & 259.7555 & 57.1735 & 119.90 &   SBc &    71.80 &    21.00 & $ 15.57_{ -5.98}^{+  5.07}$ & $  - $ & $  3.41_{ -2.02}^{+  2.10}$ & $  6.93_{ -1.49}^{+  2.27}$ & $ 10.    47_{ -1.38}^{+  2.96}$ & G \\
  manga-7992-6104 & 255.2795 & 64.6769 & 116.00 &   SBc &    46.70 &     7.90 & $ 27.36_{ -1.71}^{+  1.96}$ & $  - $ & $  0.86_{   }^{+  0.88}$ & $  2.75_{ -0.72}^{+  1.01}$ & $  5.    07_{ -0.97}^{+  1.76}$ & G \\
  manga-8082-6102 & 49.9459 & 0.5846 & 103.70 &   SB0 &    41.30 &    98.70 & $ 50.63_{-19.29}^{+ 22.90}$ & $  - $ & $  3.21_{ -0.66}^{+  0.73}$ & $  5.08_{ -1.13}^{+  1.01}$ & $  8.    13_{ -2.12}^{+  1.63}$ & G \\
 manga-8083-12704 & 50.6968 & 0.1494 &  97.70 &  SBbc &    41.70 &   144.10 & $ 84.10_{-81.25}^{+ 49.51}$ & $  - $ & $  - $ & $  1.29_{ -0.56}^{+  0.43}$ & $  2.    42_{ -0.46}^{+  0.70}$ & G \\
  manga-8133-3701 & 112.0793 & 43.3021 & 186.40 &   SBb &    44.60 &   101.20 & $ 42.71_{ -9.14}^{+  6.46}$ & $  - $ & $  1.60_{   }^{+  1.07}$ & $  3.88_{ -0.91}^{+  0.98}$ & $  6.    61_{ -1.43}^{+  1.25}$ & G \\
  manga-8134-6102 & 114.9245 & 45.9126 & 136.90 &  SB0a &    53.80 &    87.40 & $ 23.53_{ -3.92}^{+  4.85}$ & $  1.55_{ -1.15}^{+  0.84}$ & $  7.20_{ -1.59}^{+  1.95}$ & $ 12.14_{ -2.89}^{+  2.54}$ & $ 18.    67_{ -3.90}^{+  2.69}$ & G \\
  manga-8137-9102 & 117.0386 & 43.5907 & 133.10 &   SBb &    43.30 &   136.80 & $ 34.12_{ -9.04}^{+  4.52}$ & $  - $ & $  2.39_{ -0.59}^{+  0.61}$ & $  4.30_{ -0.95}^{+  1.57}$ & $  9.    00_{ -1.92}^{+  2.04}$ & G \\
 manga-8140-12701 & 116.9303 & 41.3864 & 122.40 &   SBa &    37.80 &    60.20 & $ 40.69_{ -6.32}^{+  8.52}$ & $  - $ & $  3.00_{ -0.50}^{+  0.68}$ & $  4.97_{ -0.94}^{+  1.34}$ & $  8.    80_{ -1.84}^{+  1.72}$ & G \\
 manga-8140-12703 & 117.8985 & 42.8801 & 136.90 &   SBb &    55.00 &    28.00 & $ 29.06_{ -8.09}^{+ 11.77}$ & $  - $ & $  5.23_{ -1.20}^{+  1.49}$ & $  8.87_{ -1.92}^{+  2.07}$ & $ 14.    80_{ -3.70}^{+  3.98}$ & G \\
  manga-8243-6103 & 129.1749 & 53.7272 & 134.80 &   SB0 &    59.10 &    12.10 & $ 21.93_{-15.87}^{+ 17.30}$ & $  1.11_{   }^{+  1.14}$ & $  7.28_{ -1.75}^{+  1.57}$ & $ 11.60_{ -3.33}^{+  2.16}$ & $ 17.    31_{ -4.11}^{+  2.28}$ & G \\
  manga-8244-3703 & 131.9928 & 51.6010 & 205.90 &   SB0 &    46.10 &    74.80 & $ 74.94_{-13.21}^{+ 14.60}$ & $  - $ & $  - $ & $  2.69_{ -1.54}^{+  1.07}$ & $  5.    43_{ -1.33}^{+  1.28}$ & G \\
  manga-8247-3701 & 136.6714 & 41.3651 & 107.10 &  SB0a &    37.90 &    49.70 & $ 22.89_{-11.60}^{+  5.64}$ & $  0.55_{   }^{+  0.62}$ & $  4.01_{ -1.01}^{+  1.05}$ & $  6.47_{ -1.45}^{+  1.12}$ & $  9.    58_{ -2.10}^{+  1.44}$ & G \\
  manga-8249-6101 & 137.5625 & 46.2933 & 114.30 &   SBc &    48.70 &    62.90 & $ 31.71_{ -3.36}^{+  2.88}$ & $  - $ & $  1.94_{ -0.74}^{+  0.60}$ & $  3.73_{ -0.89}^{+  1.16}$ & $  6.    70_{ -1.21}^{+  1.67}$ & G \\
  manga-8254-9101 & 161.2617 & 43.7048 & 108.40 &   SBa &    44.10 &    17.30 & $ 50.58_{-45.94}^{+ 27.07}$ & $  - $ & $  3.34_{ -0.62}^{+  0.85}$ & $  5.83_{ -1.23}^{+  1.80}$ & $ 10.    18_{ -1.81}^{+  1.91}$ & G \\
  manga-8256-6101 & 163.7348 & 41.4985 & 105.40 &   SBa &    51.40 &   132.20 & $ 37.81_{-33.30}^{+ 29.05}$ & $  - $ & $  2.76_{ -0.52}^{+  0.73}$ & $  4.93_{ -1.29}^{+  1.14}$ & $  7.    85_{ -1.63}^{+  1.38}$ & G \\
  manga-8257-3703 & 166.6557 & 46.0388 & 107.10 &   SBb &    58.30 &   156.10 & $ 51.84_{ -2.49}^{+  2.49}$ & $  - $ & $  2.74_{ -0.81}^{+  0.83}$ & $  4.78_{ -1.05}^{+  1.15}$ & $  7.    87_{ -1.79}^{+  1.51}$ & G \\
  manga-8257-6101 & 165.2613 & 44.8882 & 125.80 &   SBc &    45.00 &   159.00 & $ 49.62_{-26.67}^{+ 24.58}$ & $  - $ & $  0.75_{   }^{+  0.93}$ & $  2.83_{ -0.83}^{+  1.28}$ & $  5.    67_{ -1.14}^{+  1.88}$ & G \\
  manga-8274-6101 & 163.7348 & 41.4985 & 105.40 &   SBa &    54.00 &   129.60 & $ 15.48_{-16.45}^{+ 19.11}$ & $  1.43_{ -0.72}^{+  0.61}$ & $  6.52_{ -1.48}^{+  1.17}$ & $  9.60_{ -2.10}^{+  1.86}$ & $ 14.    73_{ -3.65}^{+  2.35}$ & G \\
 manga-8312-12702 & 245.2709 & 39.9174 & 136.90 &   SBc &    42.90 &    85.50 & $ 35.63_{ -5.76}^{+  4.87}$ & $  - $ & $  0.93_{   }^{+  0.99}$ & $  3.25_{ -0.96}^{+  1.31}$ & $  7.    41_{ -1.96}^{+  2.77}$ & G \\
 manga-8312-12704 & 247.3041 & 41.1509 & 126.70 &   SBb &    46.10 &    30.30 & $ 14.69_{ -4.52}^{+  5.20}$ & $  - $ & $  5.78_{ -1.71}^{+  1.79}$ & $ 10.13_{ -2.11}^{+  2.31}$ & $ 15.    56_{ -2.62}^{+  2.61}$ & G \\
 manga-8318-12703 & 196.2324 & 47.5036 & 167.80 &   SBb &    61.80 &    46.00 & $ 29.57_{ -8.09}^{+  6.00}$ & $  - $ & $  3.56_{ -1.15}^{+  1.06}$ & $  6.81_{ -1.69}^{+  2.99}$ & $ 15.    88_{ -4.49}^{+  4.45}$ & G \\
  manga-8320-6101 & 206.6275 & 22.7060 & 113.90 &   SBb &    50.00 &     5.90 & $ 28.37_{ -4.96}^{+  5.67}$ & $  - $ & $  3.18_{ -0.80}^{+  1.19}$ & $  6.13_{ -1.65}^{+  1.62}$ & $  9.    69_{ -1.84}^{+  1.62}$ & G \\
  manga-8326-3704 & 214.8502 & 45.9008 & 113.50 &   SBa &    50.40 &   146.10 & $ 15.57_{-40.10}^{+ 17.45}$ & $  - $ & $  3.34_{ -0.74}^{+  0.90}$ & $  5.58_{ -1.34}^{+  1.06}$ & $  8.    58_{ -2.01}^{+  1.39}$ & G \\
  manga-8326-6102 & 215.0179 & 47.1213 & 298.60 &   SBb &    51.90 &   148.00 & $ 19.40_{-13.61}^{+  8.52}$ & $  - $ & $  6.97_{ -1.43}^{+  1.63}$ & $ 11.23_{ -2.64}^{+  2.73}$ & $ 18.    03_{ -4.20}^{+  3.33}$ & G \\
 manga-8330-12703 & 203.3746 & 40.5297 & 115.20 &  SBbc &    45.00 &    75.40 & $ 46.59_{ -3.80}^{+  4.30}$ & $  - $ & $  0.36_{   }^{+  0.93}$ & $  2.10_{ -0.55}^{+  0.60}$ & $  5.    70_{ -2.44}^{+  1.82}$ & G \\
 manga-8335-12701 & 215.3953 & 40.3581 & 268.90 &   SBb &    67.00 &    81.20 & $  8.17_{ -2.75}^{+  4.58}$ & $  - $ & $ 15.41_{ -3.84}^{+  3.55}$ & $ 23.61_{ -5.03}^{+  4.30}$ & $ 33.    48_{ -6.18}^{+  5.68}$ & G \\
  manga-8439-6102 & 142.7782 & 49.0797 & 144.90 &  SBab &    49.30 &    48.90 & $ 55.20_{ -1.50}^{+  1.50}$ & $  - $ & $  1.88_{ -0.88}^{+  0.74}$ & $  3.63_{ -0.72}^{+  0.92}$ & $  6.    32_{ -1.34}^{+  1.59}$ & G \\
 manga-8439-12702 & 141.5393 & 49.3102 & 115.20 &   SBa &    55.10 &    31.30 & $ 31.87_{ -5.24}^{+  4.37}$ & $  - $ & $  3.50_{ -0.62}^{+  0.96}$ & $  6.36_{ -1.49}^{+  1.63}$ & $ 10.    99_{ -2.48}^{+  2.63}$ & G \\
 manga-8440-12704 & 136.1423 & 41.3978 & 115.60 &   SBb &    57.90 &   149.70 & $ 37.28_{ -4.42}^{+  7.79}$ & $  - $ & $  3.49_{ -0.84}^{+  0.92}$ & $  5.68_{ -1.27}^{+  1.36}$ & $  9.    24_{ -2.31}^{+  1.96}$ & G \\
  manga-8447-6101 & 206.1333 & 40.2400 & 319.00 &   SBb &    63.90 &   178.40 & $ 38.74_{-11.59}^{+  7.70}$ & $  - $ & $  4.38_{ -3.62}^{+  2.58}$ & $ 10.02_{ -2.17}^{+  2.69}$ & $ 18.    42_{ -4.48}^{+  4.28}$ & G \\
  manga-8452-3704 & 157.5390 & 47.2784 & 107.50 &   SBc &    59.70 &    72.70 & $ 79.11_{-53.11}^{+ 49.78}$ & $  - $ & $  - $ & $  - $ & $  2.    60_{ -0.89}^{+  1.08}$ & G \\
 manga-8452-12703 & 156.8057 & 48.2448 & 259.30 &   SBb &    45.70 &    75.10 & $ 43.46_{ -5.78}^{+  6.22}$ & $  - $ & $  2.90_{ -2.07}^{+  1.40}$ & $  6.10_{ -1.27}^{+  1.82}$ & $ 13.    93_{ -4.69}^{+  3.96}$ & G \\
 manga-8481-12701 & 236.7613 & 54.3409 & 284.00 &   SBa &    49.20 &   148.00 & $ 41.06_{ -7.29}^{+ 10.36}$ & $  - $ & $  3.28_{ -1.65}^{+  1.42}$ & $  6.31_{ -1.04}^{+  1.37}$ & $ 10.    71_{ -2.07}^{+  2.94}$ & G \\
  manga-8482-9102 & 242.9559 & 49.2287 & 246.70 &   SBb &    62.60 &    63.20 & $ 15.63_{ -3.86}^{+  6.03}$ & $  - $ & $  7.26_{ -1.79}^{+  2.03}$ & $ 12.27_{ -3.10}^{+  2.85}$ & $ 18.    84_{ -3.88}^{+  3.18}$ & G \\
 manga-8482-12703 & 245.5031 & 49.5208 & 211.30 &  SBbc &    42.40 &     2.90 & $ 42.42_{-16.21}^{+ 15.92}$ & $  - $ & $  2.17_{ -1.39}^{+  0.97}$ & $  4.25_{ -0.76}^{+  0.93}$ & $  7.    85_{ -1.80}^{+  3.23}$ & G \\
 manga-8482-12705 & 244.2167 & 50.2822 & 178.00 &   SBb &    63.00 &   117.20 & $ 13.14_{ -8.32}^{+  6.24}$ & $  0.88_{   }^{+  1.71}$ & $ 10.77_{ -2.77}^{+  3.09}$ & $ 17.80_{ -4.25}^{+  4.00}$ & $ 26.    79_{ -5.37}^{+  4.30}$ & G \\
  manga-8486-6101 & 238.0396 & 46.3198 & 250.50 &   SBc &    40.40 &   111.50 & $ 19.18_{ -4.83}^{+  3.94}$ & $  - $ & $  5.95_{ -1.32}^{+  2.20}$ & $  9.55_{ -1.65}^{+  3.00}$ & $ 13.    91_{ -1.68}^{+  4.01}$ & G \\
  manga-8548-6102 & 245.5224 & 46.6242 & 203.80 &   SBc &    54.10 &    64.70 & $ 35.86_{ -4.00}^{+  5.62}$ & $  - $ & $  - $ & $  3.74_{ -1.59}^{+  1.73}$ & $  7.    49_{ -1.54}^{+  2.40}$ & G \\
  manga-8548-6104 & 245.7474 & 46.6753 & 204.60 &   SBc &    62.20 &   118.10 & $ 23.82_{ -4.56}^{+  4.33}$ & $  - $ & $  4.74_{ -1.30}^{+  1.74}$ & $  7.52_{ -1.26}^{+  2.58}$ & $ 11.    27_{ -1.48}^{+  3.25}$ & G \\
 manga-8549-12702 & 241.2714 & 45.4430 & 184.80 &   SBb &    54.30 &    97.60 & $ 77.93_{-24.05}^{+ 30.92}$ & $  - $ & $  - $ & $  3.17_{ -1.00}^{+  0.91}$ & $  5.    84_{ -1.19}^{+  1.58}$ & G \\
  manga-8588-3701 & 248.1406 & 39.1310 & 545.10 &   SBb &    40.40 &   118.60 & $ 46.88_{-13.02}^{+ 13.14}$ & $  - $ & $  - $ & $  8.42_{ -2.51}^{+  2.38}$ & $ 15.    99_{ -3.80}^{+  4.55}$ & G \\
   manga-8601-12705 & 250.1231 & 39.2351 & 127.10 &   SBc &    68.30 &    49.40 & $ 23.93_{ -2.10}^{+  4.89}$ & $  - $ & $  4.10_{ -1.20}^{+  1.63}$ & $  7.70_{ -1.77}^{+  3.38}$ & $ 13.    76_{ -2.46}^{+  3.74}$ & G \\
 manga-8603-12701 & 248.1406 & 39.1310 & 545.10 &   SBb &    41.10 &   118.60 & $ 49.39_{ -9.84}^{+ 10.02}$ & $  - $ & $  - $ & $  7.84_{ -2.60}^{+  2.37}$ & $ 15.    10_{ -3.42}^{+  4.40}$ & G \\
 manga-8603-12703 & 247.2826 & 40.6650 & 128.40 &   SBa &    58.00 &    73.50 & $ 25.57_{-11.93}^{+  9.47}$ & $  - $ & $  3.07_{ -0.77}^{+  0.89}$ & $  5.73_{ -1.28}^{+  2.27}$ & $ 11.    44_{ -2.20}^{+  2.15}$ & G \\
 manga-8604-12703 & 247.7642 & 39.8385 & 130.50 &  SBab &    48.80 &   100.10 & $ 16.59_{-20.38}^{+  7.98}$ & $  1.90_{ -0.74}^{+  0.63}$ & $  7.78_{ -1.73}^{+  1.86}$ & $ 12.49_{ -3.01}^{+  3.18}$ & $ 19.    72_{ -4.23}^{+  3.60}$ & G \\
  manga-8612-6104 & 255.0069 & 38.8160 & 152.20 &   SBb &    42.40 &   169.60 & $105.51_{-13.06}^{+ 12.06}$ & $  - $ & $  - $ & $  2.11_{ -0.77}^{+  0.69}$ & $  4.    15_{ -0.91}^{+  1.17}$ & G \\
 manga-8612-12702 & 253.9464 & 39.3105 & 268.10 &   SBc &    52.30 &    49.60 & $ 41.91_{-24.02}^{+ 33.84}$ & $  - $ & $  - $ & $  1.48_{   }^{+  2.49}$ & $  7.    57_{ -2.26}^{+  3.16}$ & G \\
 manga-7990-12704 & 262.4875 & 58.3975 & 118.60 &  SBbc &    50.80 &   173.00 & $ 36.70_{ -5.00}^{+  5.70}$ & $  - $ & $  2.72_{ -0.58}^{+  0.72}$ & $  4.90_{ -1.06}^{+  1.78}$ & $  9.    55_{ -1.70}^{+  1.81}$ & Ga \\
  manga-8135-6103 & 113.0583 & 39.5600 & 201.30 &  SBab &    49.10 &    70.50 & $ 24.90_{ -3.80}^{+  1.80}$ & $  - $ & $  5.88_{ -1.19}^{+  1.59}$ & $ 10.43_{ -2.25}^{+  2.58}$ & $ 17.    19_{ -3.17}^{+  3.00}$ & Ga \\
 manga-8243-12704 & 131.1667 & 53.9511 &  99.60 &  SBbc &    51.20 &    23.30 & $ 36.00_{-11.70}^{+ 18.80}$ & $  - $ & $  - $ & $  1.91_{ -0.85}^{+  1.09}$ & $  5.    39_{ -1.62}^{+  1.94}$ & Ga \\
 manga-8341-12704 & 189.2125 & 46.6511 & 125.30 &  SBbc &    17.30 &    60.70 & $ 27.10_{ -7.70}^{+  6.70}$ & $  - $ & $  3.25_{ -0.78}^{+  1.62}$ & $  6.96_{ -1.83}^{+  1.74}$ & $ 11.    52_{ -2.64}^{+  1.97}$ & Ga \\
 manga-8453-12701 & 151.3083 & 46.6508 & 103.70 &  SABc &    36.90 &    98.50 & $ 28.30_{ -3.00}^{+ 15.10}$ & $  - $ & $  1.50_{ -0.50}^{+  0.41}$ & $  2.76_{ -0.65}^{+  1.45}$ & $  6.    87_{ -1.64}^{+  2.13}$ & Ga \\
 \hline
\enddata
\tablecomments{(1) Galaxy ID (2) Right ascension (3) Declination (4) Distance (5) Hubble type (6) Inclination (7) Position angle (8) Bar pattern speed measured by the TW method. They are come from the reference paper (13): A, C, G, and Ga stand for \citet{2015Aguerri}, \citet{2019Cuomo}, \citet{2019Guo}, and \citet{2020Garma-Oehmichen}. $R_{\rm ILR}$ (9), $R_{\rm UHR}$ (10), $R_{\rm CR}$ (11), and $R_{\rm OLR}$ (12) are the locations of ILR, UHR, CR, and OLR we measure where the bar pattern speed intersects the frequency curves.}
\end{deluxetable*}
\end{longrotatetable}



\end{document}